\documentclass[%
reprint,
groupedaddress,
 amsmath,amssymb,
 aps,
 pra,
]{revtex4-2}

\usepackage{graphicx}
\usepackage{dcolumn}
\usepackage{bm}
\usepackage{textcomp}

\usepackage{appendix}

\begin{document}

\preprint{APS/123-QED}

\title{Visualizing Detection Efficiency in Optomechanical Scattering
}

\author{Youssef Tawfik}
\affiliation{Department of Physics and Astronomy, University of Pittsburgh, Pittsburgh, PA, USA}

\author{Shan Hao}%
\altaffiliation[Present address: ]{Apple. Inc}
\affiliation{Department of Physics and Astronomy, University of Pittsburgh, Pittsburgh, PA, USA}

\author{Thomas P. Purdy}%
 \email{Contact Author: tpp9@pitt.edu}
\affiliation{Department of Physics and Astronomy, University of Pittsburgh, Pittsburgh, PA, USA}
\date{\today}

\begin{abstract}
Many optical measurement techniques, such as light scattering from wavelength-scale particles or detecting motion from a surface with an optical lever, encode information in a complex radiation pattern. Extracting all available information is essential for many quantum-enhanced sensing protocols but is often impractical, as it requires many channels to spatially resolve the scattered signal. We present a new method to visualize how efficiently a practical measurement scheme captures the information available in the scattered light by mapping out the local contribution to the detection efficiency on the detector surface. We use this tool to experimentally optimize the free space measurement of the amplitude of motion of an optomechanical resonator with a quadrant photodiode. We show that blocking sections of the photodetector enhances sensitivity, counterintuitively yielding a significant improvement in detecting higher-order mechanical modes in the system. We also show how our method can be applied to light scattering measurements of small particles.
\end{abstract}

\maketitle

 Quantum enhanced sensors tailor states of light and measurement protocols to reduce noise beyond what can be achieved classically \cite{Defienne_Bowen_Chekhova_Lemos_Oron_Ramelow_Treps_Faccio_2024,Pirandola_Bardhan_Gehring_Weedbrook_Lloyd_2018}. Most quantum sensing protocols require efficient detection to take full advantage of quantum resources \cite{Lawrie_Lett_Marino_Pooser_2019}. In some applications such as interferometric detection of gravitational waves, information is encoded in a single channel of light, and the detection schemes and quantum states to achieve high efficiency and suppress noise are well understood \cite{Ganapathy2023,Whittle2021}. However, many applications --- such as light scattering from wavelength-scale particles \cite{Tebbenjohanns_Frimmer_Novotny_2019,Gonzalez-Ballestero2021} or optical lever detection   \cite{Hao_Purdy_2024,Pluchar_He_Manley_Deshler_Guha_Wilson_2025,Pratt_Agrawal_Condos_Pluchar_Schlamminger_Wilson_2023,Tay2008} --- encode information in spatially complex radiation patterns. Extracting all available information is often impractical, as it requires precise knowledge of the system geometry and many measurement channels to spatially resolve the scattered signal. To gain quantum-enhanced sensitivity in these messy free-space measurements, practical detection techniques that are close to optimal must be devised.
 
We present a new method to characterize how efficiently free-space measurement schemes capture the information available in the scattered light, which we dub the differential detection efficiency. We use this tool to optimize the free-space measurement of the modal amplitude of motion of a SiN membrane over measurement schemes including quadrant photodetection.  We experimentally demonstrate that blocking selected regions of a quadrant photodetector (QPD) enhances sensitivity, yielding a several-fold improvement in detecting higher-order mechanical modes of the membrane.  We also show how our methods can be applied to optimize position sensing of small particles in the Rayleigh scattering limit.  Free-space measurements such as quadrant photodetection offer practical advantages over single-mode interferometery: they are simple to implement, are insensitive to path-length and laser phase noise, often do not require precise spatial mode matching, and have comparable sensitivity to interferometery \cite{Putman_de_Grooth_van_Hulst_Greve_1992}.  However, their quantum limits and optimality have been much more difficult to characterize. 

An optical displacement measurement is characterized by two fundamental noise contributions, both arising from the shot noise fluctuations on the light, the imprecision noise floor, $S_X^{\text{imp}}$, due to directly detected shot noise and the back action force noise, $S_F^{\text{ba}}$, due to the noisy optical force imparted by the light scattering from the system. These quantities are subject to a Heisenberg uncertainty relation, $S_X^{\text{imp}}S_F^{\text{ba}}\geq (\hbar/2)^2$ \cite{Aspelmeyer_Kippenberg_Marquardt_2014, Clerk_Devoret_Girvin_Marquardt_Schoelkopf_2010}.  For a given input beam, the available position information carried by the scattered light is fixed and is characterized by an ideal measurement imprecision $S_X^\text{ideal}=\hbar^2/4 S_F^\text{ba}$.   An ideal measurement scheme can be generated by interfering the scattered light with an optimized, spatially dependent, homodyne local oscillator that weights each direction based on the local sensitivity and noise~\cite{Tebbenjohanns_Frimmer_Novotny_2019}.   The information contained in the scattered light and extracted by this ideal measurement scheme can be visualized by recently developed quantity, the information radiation pattern (IRP)~\cite{Tebbenjohanns_Frimmer_Novotny_2019,Maurer_Gonzalez-Ballestero_Romero-Isart_2023,Hupfl_Russo_Rachbauer_Bouchet_Lu_Kuhl_Rotter_2024}. However, the ideal protocol is typically impractical to implement, and realistic measurement schemes capture only a fraction of the information, called the detection efficiency $\eta=S_X^\text{ideal}/S_X^\text{imp}$. 

\begin{figure}  
  \centering
  \includegraphics{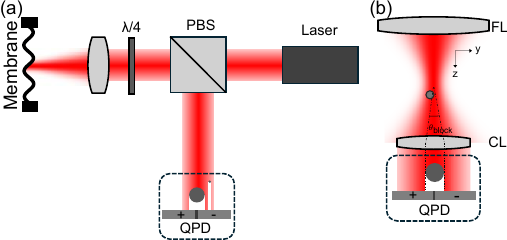}
  \caption{Examples of free space optomechanical scattering experiments. (a) Detection of the vibrational modes of membrane optomechanical resonator via optical lever detection. Setup includes a polarizing beam splitter (PBS), quarter wave plate ($\lambda/4$), and quadrant photodetector (QPD). (b)Detection of the location of a small particle via light scattering using high numerical aperture focusing lens (FL) and collection lens (CL).  In either case information is scattered into many optical modes which are analyzed with multiple detectors which may be partially blocked assess and optimize the detection efficiency.}
  \label{fig:cartoon}
\end{figure}

We consider measurements where light scatters from a target object, propagates through an optical system, and is detected by an array of photodetectors (Fig.~\ref{fig:cartoon}). Information can be lost at each step in this process, reducing the net detection efficiency.  Light is absorbed or lost to finite numerical apertures along the optical path, limiting the collection efficiency, $\eta_{\text{col}}$; the photodetectors have finite conversion efficiency, $\eta_{\text{qe}}$; some of the information is encoded in a quadrature orthogonal to that being measured.  The main focus of this paper is on visualizing and optimizing an additional geometric detection efficiency where information is lost when the geometry of a local oscillator or the spatial weighting of the  photodetected signals is not ideal.


Our goal is to measure a small parameter $A$ of the system.  For concreteness, we assume this parameter to be a spatial degree of freedom, such as the location~\cite{Tebbenjohanns_Frimmer_Novotny_2019,Dania_Heidegger_Bykov_Cerchiari_Araneda_Northup_2022,Maurer_Gonzalez-Ballestero_Romero-Isart_2023,Tebbenjohanns2021,Magrini2021} or orientation of a levitated particle~\cite{Tebbenjohanns2022,Laing_Klomp_Winstone_Grinin_Dana_Wang_Widyatmodjo_Bateman_Geraci_2024, Gao_van_der_Laan_Zielińska_Militaru_Novotny_Frimmer_2024,Bang2020} or the amplitude of a vibrational mode~\cite{Pratt_Agrawal_Condos_Pluchar_Schlamminger_Wilson_2023,Hao_Purdy_2024,Rossi2018,Shin2025}.  An input beam in spatial mode $u^{\text{in}}\left(\boldsymbol{r}\right)$, with wavenumber  $k$, and amplitude $\alpha$, strikes the system and is scattered or reflected into an output mode, $u^{\text{out}}\left(\boldsymbol{r}\right)$. For $kA\ll1$ this mode can be expanded to first order,
\begin{equation}
u^{\text{out}}\left(\boldsymbol{r}\right)=u_{0}^{\text{out}}\left(\boldsymbol{r}\right)+kA u_{\text{s }}^{\text{out}}\left(\boldsymbol{r}\right)
\end{equation}
containing a stationary part, $u^{\text{out}}_0$, independent of $A$ and a signal field, $u^{\text{out}}_{\text{s }}=\frac{1}{k}du^{\text{out}}/dA$,  containing information about $A$.  The intensity at the detection surface, $D$, is $I\left(\boldsymbol{r}\right)=\left|  \alpha \,  u^{\text{out}}\left(\boldsymbol{r}\right)\right|^2$.   We model the detector by a weight function $f_w\left(\boldsymbol{r}\right)$ defined over all of $D$ (e.g.~$f_w= \pm 1$ on the opposite halves of a QPD and $f_w=0$ outside the detector). The detected signal is then
\begin{equation}
V=\int_{D}f_w\left(\boldsymbol{r}\right) I\left(\boldsymbol{r}\right)  da
\end{equation}
We define a measurement sensitivity $\mathcal{S}=dV/dA=2\alpha^2 k\int_{D}f_w\text{Re}\left[ u_{0}^{\text{out}}u_{\text{s}}^{\text{out}*}\right]da$, such that $V/\mathcal{S}$ is our estimate of $A$.  We assume that the noise on our measurement is dominated by the shot noise fluctuations on $u^{\text{out}}_0$ weighted by $f_w$, and that the noise on each photodetector is uncorrelated, leading to the definition $\mathcal{N}=\int_{D}\left|\alpha  u^{\text{out}}_0\left(\boldsymbol{r}\right)\right|^2f_w\left(\boldsymbol{r}\right)^2 da$.  As shown in Appendix~\ref{AppendixA}, the measurement imprecision noise spectral density is given by $S_A^{\text{imp}}=\mathcal{N}/\mathcal{S}^2$ and the detection efficiency becomes $\eta=S_A^{\text{ideal}}\frac{\mathcal{S}^2}{\mathcal{N}}$ , where $S_A^{\text{ideal}}=\left(4\alpha^2 k^2\mathcal{}\int_{D}\left| u_{\text{s}}^{\text{out}}\left(\boldsymbol{r}\right)\right|^2da\right)^{-1}$.  Then the back action force noise takes a particularly simple and general form
\begin{equation}\label{eq:Sbamain}
S^{\text{ba}}_F=\alpha^2 \hbar^2\int_D\left| \frac{du^{\text{out}}}{dA}\right|^2da
\end{equation}
The ideal measurement imprecision and back action noise can also be obtained by considering the quantum Fisher information~\cite{Bouchet_Rotter_Mosk_2021,Hupfl_Russo_Rachbauer_Bouchet_Lu_Kuhl_Rotter_2024} (See supplemental materials~\cite{supp}).  

\begin{figure}  
  \centering
  \includegraphics{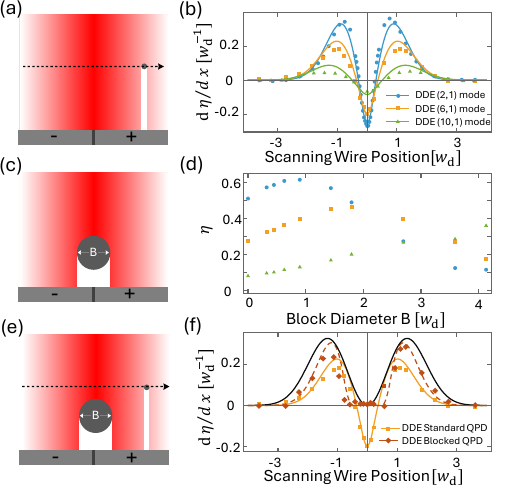}
  \caption{Differential Detection Efficiency (DDE).  (a,b) A thin wire is scanned in front of the QPD to locally exclude light and measure the DDE ($d\eta/dx$).  Theoretical (solid curves) and Measured (data points) DDE for several membrane modes are displayed. (c,d) A beam block, diameter $B$, masks the center of the QPD, while detection efficiencies for several membrane modes (symbols are the same as in (b)) are measured. (e,f) The DDE of the partially blocked QPD (red) approaches the IRP (Black) for the (6,1) membrane mode.  For all data, statical errors are smaller than the size of the symbols.}
  \label{fig:detadx}
\end{figure}

We next construct a new quantity, the differential detection efficiency $d\eta/da$, which is a function that maps out the contribution of each point in the detection plane to the overall detection efficiency.   We compute the differential detection efficiency (DDE) at each location $\boldsymbol{r}$ by evaluating the change in efficiency, $d\eta$, between including and excluding an area element $da$ located at $\boldsymbol{r}$ in the overall detection scheme.  We find (see Appendix~\ref{AppendixA}) 
\begin{align}\label{eq:dde}
\frac{d\eta}{da} =\alpha^2 S_{A}^{\text{ideal}}
&\left(4k\left(\frac{\mathcal{S}}{\mathcal{N}}\right) \text{Re}\left[u_{0}^{\text{out}}(\boldsymbol{r}) u_{\text{s}}^{\text{out}*}(\boldsymbol{r}) f_w(\boldsymbol{r})\right]\right.\\
  & \left.
-\left(\frac{\mathcal{S}}{\mathcal{N}}\right)^2 \left|u_{0}^{\text{out}}(\boldsymbol{r}) f_w(\boldsymbol{r})\right|^{2}\right), \nonumber
\end{align}
Integrating the DDE over the detection plane yields the detection efficiency, $\int_D \frac{d\eta}{da}da=\eta$.   The first term in Eq.~\ref{eq:dde} represents the contribution to the signal at $\boldsymbol{r}$, and the second term represents a penalty from the shot noise at that location.   Importantly, the DDE can be negative over certain regions of $D$ (See Fig.~\ref{fig:detadx}), meaning these regions contain very little signal but high levels of noise and are relatively overweighted by $f_w$.  In this case, the efficiency (and signal to noise ratio) of a detection scheme can be improved by excluding (i.~e. simply blocking) these regions of the detector.  An ideal DDE can be computed, based on the ideal spatially resolved homodyne detector, which corresponds to the IRP when the detection plane is in the far field~\cite{Tebbenjohanns_Frimmer_Novotny_2019}, $\left(\frac{d\eta}{da}\right)^\text{ideal}=4\alpha^2 k^2S_{A}^{\text{ideal}}\left|u_{\text{s}}^{\text{out}}\left(\boldsymbol{r}\right)\right|^2$.  By comparing the actual DDE to the ideal one, we can visualize the local efficiency of a detection scheme and devise methods for improvement.

To showcase the utility of our approach, we optically probe the motion of the flexural modes of a rectangular, low-stress silicon nitride membrane of dimensions 1.5~mm $\times$ 3.5~mm $\times$ 100~nm (Norcada Inc. NX53515C).   We focus a Gaussian beam from a 1064~nm laser normally incident on the center ($x=y=0$) of the membrane and detect the reflected light on a QPD in the far field (See Fig.~\ref{fig:cartoon}(a)).  The surface displacement of the membrane is $\Psi(x,y, t) = \sum_{mn} A_{mn}(t) \psi_{mn}(x,y)$, where $\psi_{mn}(x,y)$ are the sinusoidal mode functions of the membrane, and $m$ and $n$ are the number of antinodes along the $x$ and $y$ axes of the membrane, respectively.  The unknown parameters to be measured are the modal amplitudes $A_{mn}$.  

We analyze the signal from the QPD by taking the difference between the photocurrents on the right and left halves of the detector.  This signal is sensitive to modes where $m$  is even and $n$ is odd (i.~e.~our laser is centered horizontally on a node line and vertically on an antinode).  Qualitatively, when the beam waist, $w_0$ is much smaller than the mechanical wavelength, ($k_m w_0 \ll 1$, where $k_m$ is the mechanical wavenumber in the $x$ direction) our light is effectively reflecting from a flat tilting surface, equivalent to an optical lever.  In this case, we can expand the output field in the basis of Hermite Gaussian functions, $HG_{i,j}\left(\boldsymbol{r}\right)$, and find $u_{\text{s}}^{\text{out}}\propto HG_{1,0}$~\cite{Enomoto2016}.  When $k_m w_0 \gg 1$, the surface looks like a diffraction grating, and $u_{\text{s}}^{\text{out}}$ takes the form of two first order diffraction peaks with an angular separation $\sim \frac{k_m}{k}$.  Quantitatively, motion of the surface creates a path length change corresponding to a phase shift of $2k \Psi(x,y, t)$ on the input beam.  Then the optical field immediately after reflection is $u^{\text{in}}e^{i2k \Psi}$.  This field can then be propagated to the detection plane and its intensity integrated over the photodetector elements to arrive at the expected signal~\cite{supp}.

\begin{figure}  
  \centering
  \includegraphics{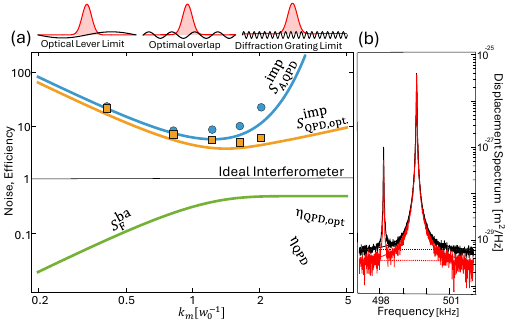}
  \caption{
  Measurement Sensitivity relative to an ideal single mode interferometer. (a) Imprecision noise and efficiency for membrane modes of various $k_m$ measured with a standard QPD (Blue) or an optimally blocked QPD (Yellow), and the back action noise (Green). All values are reported relative to those of a single mode interferometer \cite{supp}. Top diagrams show relative size of Gaussian beams (Red) and the mechanical wavelength (Black)  (b)  Thermal displacement spectrum of the (8,1) mode of the membrane measured using an standard QPD (Red) and an optimally gaped QPD (Black) where dashed horizontal lines represent the shot noise floor.
  }
  \label{fig:system_limits}
\end{figure}

The spectrum of the measured signal takes the form of a series of Lorentzian peaks, representing the thermal motion of several mechanical modes, on top of a white shot noise floor, representing the measurement imprecision noise, $S_{A_{mn}}^{\text{imp}}$ (See Fig.~\ref{fig:system_limits}b).  Near each mechanical resonance, we calibrate the spectrum, assuming that the mode is thermally occupied at room temperature, so that the area under the peak of a mode is $\left\langle x^{2}\right\rangle =k_B T/m_{\text{eff}}\omega_{mn}^2$, where $\omega_{mn}$ is the mode frequency and $m_{\text{eff}}$ is mode effective mass.  We estimate $m_{\text{eff}}=390\pm20$~ng with the uncertainty limited by our knowledge of the film density.  This calibration leaves the shot noise floor in units of displacement noise spectral density.  We confirm that the noise floor scales linearly with optical power in a frequency band near each mode, the expected signature of optical shot noise~\cite{supp}. 

We can estimate the back action force noise and ideal measurement imprecision from the optical and mechanical mode profiles.  The integral in Eq.~\ref{eq:Sbamain} can be evaluated over any surface through which all of the light passes. Immediately after reflection, at the device plane, $S^{\text{ba}}_{F_{mn}}=4\alpha^2\left(\hbar k\right)^2\int \left|u^{\text{in}}\right|^2\psi_{mn}^2da$, which can be interpreted as an incoherent sum of the local radiation pressure noise fluctuations over the surface of the membrane~\cite{Pinard1999, Hao_Purdy_2024, Pluchar_He_Manley_Deshler_Guha_Wilson_2025}.  The leading uncertainties come from the  beam waist, $w_{0}=100\pm5$~\textmu m, estimated from the geometrical parameters of our optical input beam path, and the 5\% uncertainty of our optical power meter.  Comparing the measured noise floor to the ideal one, we find $\eta$ for each mechanical mode.  For the (2,1) mode, we measure $\eta=0.504\pm0.06$.  Since the measurement of this mode is well approximated as an optical lever detector ($k_m w_0\ll1$), the expected efficiency simplifies to $\eta=\eta_{\text{qe}}\frac{2}{\pi}=0.55$, in agreement with the measured value.  Here, $\eta_{\text{qe}}=0.87$ is the measured photodetector conversion efficiency.   The factor $2/\pi$ is the geometric efficiency for measuring the center of a Gaussian beam on a QPD~\cite{Knee2015,Walborn_Aguilar_Saldanha_Davidovich_Filho_2020,Hsu_Delaubert_Lam_Bowen_2004,Pluchar_He_Manley_Deshler_Guha_Wilson_2025, Hao_Purdy_2024}.

Measurements of higher order mechanical modes (larger $m$) qualitatively show the expected trend towards lower $\eta$, but quantitatively disagree with predictions.  We attribute this discrepancy to aberrations in the Gaussian wings of the probe beam, which increasingly affect the overlap integral in $\mathcal{S}$ as we move into the diffraction grating limit.  This motivates a more careful, empirical investigation of the spatial distribution of the optically encoded information.  Our definition of the DDE lends itself to a simple experimental measurement procedure.  With our measurement running as previously described, we can sequentially block small areas of the photodetector, $\Delta a$, and record the change in efficiency, $\Delta \eta$.  The DDE is estimated as the finite difference, $\Delta \eta/\Delta a$.  We can treat our particular case of probing $n=1$ modes as effectively 1-D~\cite{supp} and block narrow strips of the detector, measuring $\Delta\eta/\Delta x$.  To locally block light, we insert a thin, vertically oriented wire, (diameter $\Delta x$=180 \textmu m) a few mm in front of the QDP, and control its lateral position with a motorized translation stage (See Fig.~\ref{fig:detadx}(a)).  We choose $\Delta x$ to be smaller than $w_d$=560~\textmu m, the beam waist at the detector and the 3~mm diameter of the QPD.  To make faster and more precise measurements of the relative efficiency, we drive the membrane via a piezoelectric shaker with a coherent force near its resonance, and measure the signal amplitude at the drive frequency.  We compare this peak to the average power on the photodetector (which scales linearly with the shot noise floor) to estimate the SNR.  Paired with the previous thermal motion calibrations, this procedure allows us to rapidly map out the DDE with low statistical uncertainty. 

Figure~\ref{fig:detadx}(b) shows the measured and predicted DDEs for several mechanical modes.  A key feature is a region near the center of the QPD where $d\eta/dx$ is negative.  Here, $u^{\text{out}}_0$ is maximal, meaning a large shot noise contribution and $u^{\text{out}}_{s}$ is small, while this region is weighted the same  ($\left|f_w\right|=1$ everywhere) as more information-dense regions away from the center.  This observation leads to a simple procedure to improve our photodetection setup -- physically blocking the center of the QPD.  As shown in Fig.~\ref{fig:detadx}(c), we insert a long, vertically oriented block of width $B$ centered in front of the QPD, in our case a blackened steel rod of diameter $B$.  We find that $\eta$ increases for small $B$ and is maximized when a substantial fraction of the total reflected power is blocked.  The optimum block size increases with the mode index $m$, consistent with the notion that  $u^{\text{out}}_{\text{s}}$ peaks at larger $|x|$  as  $k_m$ increases.  Remarkably, for the (10,1) mode, we find that blocking more than 95\% of the reflected power increases $\eta$ and the SNR by over a factor of 3.

We remeasure the DDE with a partially block QPD, as shown in Fig.~\ref{fig:detadx}(e,f).  With this optimized detection scheme, the DDE is no longer negative anywhere and its value is closer to the envelope of an ideal detection scheme.  We note that the optimization is not exactly equivalent to blocking all of the negative areas of the DDE for the standard QPD.  As light is blocked, both $\mathcal{S}$ and $\mathcal{N}$ change, shifting the threshold for when the DDE is positive or negative.  We can also understand our optimization procedure as process for (crudely) modifying $f_w$.  For example, in the optical lever limit, choosing a linear weighting function $f_w\propto x$ yields $\eta=1$~\cite{Fradgley_French_Rushton_Dieudonné_Harrison_Beckey_Miao_Gill_Petrov_Boyer_2022}.  So, in effect, the step function response of a standard QPD is a poor approximation of a line, whereas the staircase-like weighting function introduced by the block is a better approximation~\cite{supp}.

The results of our investigation of detecting membrane motion are summarized in Fig.~\ref{fig:system_limits}.  Here, we compare the measured and expected sensitivity to that of a single-mode interferometer with the same reflected power~\cite{Putman_de_Grooth_van_Hulst_Greve_1992} (e.~g.~an unbalanced Michelson interferometer, where in one arm, the beam is tightly focused onto an antinode of the mechanical mode).  In the optical lever limit, the sensitivity increases with $k_m$ as the slope of the membrane increases.  The beam samples an increasingly large region where the mechanical mode has a large amplitude, generating a larger back action force noise, leaving $S_{A_{mn}}^{\text{imp}} S_{F_{mn}}^{\text{ba}}$ and $\eta$ constant. In the regime where $k_m w_0\sim1$, the measurement sensitivity is at its best and can approach within a factor of a few of ideal single-mode interferometry.  At large $k_m$, standard QPD detection rapidly loses its sensitivity, as the scattering into first order diffraction spots become well resolved from the (zero-order) reflected Gaussian beam.  Much of this sensitivity can be regained by blocking this zero-order peak which contains mostly noise.

Our framework of efficiency visualization and optimization can be applied to a wide range of commonly used optical detection methods.  Techniques such as knife-edge displacement detection, resolving signals with optical mode sorters~\cite{Dinter_Roberts_Volz_Schmidt_Laplane_2024,Choi_Pluchar_He_Guha_Wilson_2024}, or optical speckle sensing~\cite{Chapalo2024,Lengenfelder2021} are amenable to improvement.  In the supplemental materials~\cite{supp}, we analyze phase contrast imaging~\cite{hecht} to detect the motion of a membrane.  It is natural to consider such a technique that maps phase information in the device plane to amplitude information in an image plane.  While a high resolution camera in the image plane could, in principle, yield high detection efficiency, for fast signals and substantial optical power, most cameras lack the frame rate and dynamic range to realize efficient detection.  We find that high efficiency detection can be obtained with a careful arrangement of only a minimal number of photodetectors.

 \begin{figure*}  
  \centering
  \includegraphics{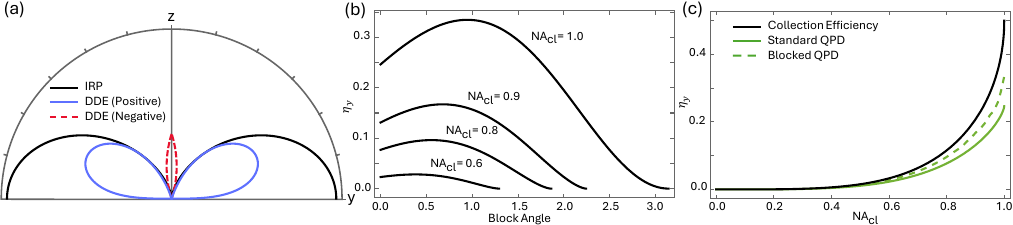}
  \caption{
  Detection efficiency of the $y$ direction motion of dipolar scatterer. (a) Polar plot of the $x=0$ cross section of the far-field  DDE (blue) with negative regions represented by dashed red line for  $\text{NA}_\text{cl}=1$ and the IRP (black).  Note, the IRP is symmetric about the $y$ axis, while the DDE is zero in the backwards directions.   The full 3-D distribution is presented in the supplemental materials~\cite{supp}. (b) The detection efficiency $\eta_y$ as a function of the amount of blocked light, for different collection lens NA. (c) Detection efficiency $\eta_y$ as a function of the with a block angle chosen to maximize $\eta_y$.
}
  \label{fig:dipolar}
\end{figure*}

We next consider the task of tracking the trajectory of a small dielectric particle via light scattering with high speed and high sensitivity~\cite{Kheifets2014}.  This task forms the basis of levitated quantum optomechanics~\cite{Gonzalez-Ballestero2021}, where high efficiency position measurement is vital for preparing particles in their motional ground state~\cite{Tebbenjohanns2021,Magrini2021} and for observing the ponderomotive optical correlations that generate optical squeezed states~\cite{Militaru2022,Margrini2022} for position detection below the standard quantum limit.   For a subwavelength particle, in the Rayleigh scattering limit, probed with laser focused through a high numerical aperture (NA) lens, the output light consists of the unperturbed diverging laser and the dipole radiation pattern of the particle (Fig.~\ref{fig:cartoon}(b)).    A fraction of the output light is collected through another high NA objective lens.    Following Ref.~\cite{Tebbenjohanns_Frimmer_Novotny_2019}, we can compute the ideal imprecision and the ideal DDE for measuring a particular  coordinate of the particle~\cite{supp}.  In Fig.~\ref{fig:dipolar}(b,c) we assess the performance of using a QPD to monitor the position of the particle in the $y$-direction, transverse to the optical axis and orthogonal to the polarization direction. 
As is the case for measuring membrane motion, the DDE shows a negative lobe near the center of the QPD.  Blocking this region of the detector generally improves $\eta$.  Unlike the case of the membrane, information about the particle position is scattered over a wide range of directions.  Capturing only a fraction of this signal over the area of the collection lens, $D_c$, limits the information that can be extracted by the QPD.  We can then break up $\eta=\eta_{\text{col}}\eta_{\text{QPD}}$ into two pieces: $\eta_{\text{col}}=\int_{D_c}\left(\frac{d\eta}{d\Omega}\right)^{\text{ideal}}d\Omega$, which is the fraction of the information collected by the lens, and $\eta_{\text{QPD}}$, the fraction of the information extracted by the QPD from the light that makes it through the lens.  As an example, for detecting motion along the $y$ direction and assuming NA$=1$ lenses, $\eta_{col}=0.5$ and 
 $\eta$ increases from 0.25 with a standard QPD to 0.34 for an optimized partially blocked detector~\cite{supp}. Considering larger, few-wavelength scale particles, in the Mie scattering limit, a larger fraction of the position information is scattered into forward directions~\cite{Maurer_Gonzalez-Ballestero_Romero-Isart_2023}, opening up the possibility of high efficiency detection of more massive particles.

In summary, we have developed a theoretical and empirical framework to visualize and optimize free-space optical sensors.  We have demonstrated that it is not always best to simply collect as much light as possible on a photodetection system.  We have also demonstrated that simple, robust, and common measurement techniques, such as quadrant photodetection, are able to approach the sensitivity and efficiency of traditional single-mode interferometry, and should be considered as viable options for sensing applications where quantum limits matter.  Quantum sensing protocols such as active feedback cooling of mechanical systems into their ground state~\cite{Tebbenjohanns2021,Magrini2021,Rossi2018}, or probing systems with squeezed light~\cite{Treps2003, Pooser2015}, require threshold levels of efficiency.  Our work shows that geometric contributions to the detection efficiency cannot be overlooked in such applications.

\begin{acknowledgments}
This research was supported by the National Science Foundation Grant no.~2047823.  This project was made possible through the support of Grant 63121 from the John Templeton Foundation.
\end{acknowledgments}

\bibliography{apssamp}

\begin{thebibliography}{44}%
\makeatletter
\providecommand \@ifxundefined [1]{%
 \@ifx{#1\undefined}
}%
\providecommand \@ifnum [1]{%
 \ifnum #1\expandafter \@firstoftwo
 \else \expandafter \@secondoftwo
 \fi
}%
\providecommand \@ifx [1]{%
 \ifx #1\expandafter \@firstoftwo
 \else \expandafter \@secondoftwo
 \fi
}%
\providecommand \natexlab [1]{#1}%
\providecommand \enquote  [1]{``#1''}%
\providecommand \bibnamefont  [1]{#1}%
\providecommand \bibfnamefont [1]{#1}%
\providecommand \citenamefont [1]{#1}%
\providecommand \href@noop [0]{\@secondoftwo}%
\providecommand \href [0]{\begingroup \@sanitize@url \@href}%
\providecommand \@href[1]{\@@startlink{#1}\@@href}%
\providecommand \@@href[1]{\endgroup#1\@@endlink}%
\providecommand \@sanitize@url [0]{\catcode `\\12\catcode `\$12\catcode `\&12\catcode `\#12\catcode `\^12\catcode `\_12\catcode `\%12\relax}%
\providecommand \@@startlink[1]{}%
\providecommand \@@endlink[0]{}%
\providecommand \url  [0]{\begingroup\@sanitize@url \@url }%
\providecommand \@url [1]{\endgroup\@href {#1}{\urlprefix }}%
\providecommand \urlprefix  [0]{URL }%
\providecommand \Eprint [0]{\href }%
\providecommand \doibase [0]{https://doi.org/}%
\providecommand \selectlanguage [0]{\@gobble}%
\providecommand \bibinfo  [0]{\@secondoftwo}%
\providecommand \bibfield  [0]{\@secondoftwo}%
\providecommand \translation [1]{[#1]}%
\providecommand \BibitemOpen [0]{}%
\providecommand \bibitemStop [0]{}%
\providecommand \bibitemNoStop [0]{.\EOS\space}%
\providecommand \EOS [0]{\spacefactor3000\relax}%
\providecommand \BibitemShut  [1]{\csname bibitem#1\endcsname}%
\let\auto@bib@innerbib\@empty
\bibitem [{\citenamefont {Defienne}\ \emph {et~al.}(2024)\citenamefont {Defienne}, \citenamefont {Bowen}, \citenamefont {Chekhova}, \citenamefont {Lemos}, \citenamefont {Oron}, \citenamefont {Ramelow}, \citenamefont {Treps},\ and\ \citenamefont {Faccio}}]{Defienne_Bowen_Chekhova_Lemos_Oron_Ramelow_Treps_Faccio_2024}%
  \BibitemOpen
  \bibfield  {author} {\bibinfo {author} {\bibfnamefont {H.}~\bibnamefont {Defienne}}, \bibinfo {author} {\bibfnamefont {W.~P.}\ \bibnamefont {Bowen}}, \bibinfo {author} {\bibfnamefont {M.}~\bibnamefont {Chekhova}}, \bibinfo {author} {\bibfnamefont {G.~B.}\ \bibnamefont {Lemos}}, \bibinfo {author} {\bibfnamefont {D.}~\bibnamefont {Oron}}, \bibinfo {author} {\bibfnamefont {S.}~\bibnamefont {Ramelow}}, \bibinfo {author} {\bibfnamefont {N.}~\bibnamefont {Treps}},\ and\ \bibinfo {author} {\bibfnamefont {D.}~\bibnamefont {Faccio}},\ }\bibfield  {title} {\bibinfo {title} {Advances in quantum imaging},\ }\href {https://doi.org/10.1038/s41566-024-01516-w} {\bibfield  {journal} {\bibinfo  {journal} {Nature Photonics}\ }\textbf {\bibinfo {volume} {18}},\ \bibinfo {pages} {1024–1036} (\bibinfo {year} {2024})}\BibitemShut {NoStop}%
\bibitem [{\citenamefont {Pirandola}\ \emph {et~al.}(2018)\citenamefont {Pirandola}, \citenamefont {Bardhan}, \citenamefont {Gehring}, \citenamefont {Weedbrook},\ and\ \citenamefont {Lloyd}}]{Pirandola_Bardhan_Gehring_Weedbrook_Lloyd_2018}%
  \BibitemOpen
  \bibfield  {author} {\bibinfo {author} {\bibfnamefont {S.}~\bibnamefont {Pirandola}}, \bibinfo {author} {\bibfnamefont {B.~R.}\ \bibnamefont {Bardhan}}, \bibinfo {author} {\bibfnamefont {T.}~\bibnamefont {Gehring}}, \bibinfo {author} {\bibfnamefont {C.}~\bibnamefont {Weedbrook}},\ and\ \bibinfo {author} {\bibfnamefont {S.}~\bibnamefont {Lloyd}},\ }\bibfield  {title} {\bibinfo {title} {Advances in photonic quantum sensing},\ }\href {https://doi.org/10.1038/s41566-018-0301-6} {\bibfield  {journal} {\bibinfo  {journal} {Nature Photonics}\ }\textbf {\bibinfo {volume} {12}},\ \bibinfo {pages} {724–733} (\bibinfo {year} {2018})}\BibitemShut {NoStop}%
\bibitem [{\citenamefont {Lawrie}\ \emph {et~al.}(2019)\citenamefont {Lawrie}, \citenamefont {Lett}, \citenamefont {Marino},\ and\ \citenamefont {Pooser}}]{Lawrie_Lett_Marino_Pooser_2019}%
  \BibitemOpen
  \bibfield  {author} {\bibinfo {author} {\bibfnamefont {B.~J.}\ \bibnamefont {Lawrie}}, \bibinfo {author} {\bibfnamefont {P.~D.}\ \bibnamefont {Lett}}, \bibinfo {author} {\bibfnamefont {A.~M.}\ \bibnamefont {Marino}},\ and\ \bibinfo {author} {\bibfnamefont {R.~C.}\ \bibnamefont {Pooser}},\ }\bibfield  {title} {\bibinfo {title} {Quantum sensing with squeezed light},\ }\href {https://doi.org/10.1021/acsphotonics.9b00250} {\bibfield  {journal} {\bibinfo  {journal} {ACS Photonics}\ }\textbf {\bibinfo {volume} {6}},\ \bibinfo {pages} {1307–1318} (\bibinfo {year} {2019})}\BibitemShut {NoStop}%
\bibitem [{\citenamefont {Ganapathy}\ \emph {et~al.}(2023)\citenamefont {Ganapathy} \emph {et~al.}}]{Ganapathy2023}%
  \BibitemOpen
  \bibfield  {author} {\bibinfo {author} {\bibfnamefont {D.}~\bibnamefont {Ganapathy}} \emph {et~al.} (\bibinfo {collaboration} {LIGO O4 Detector Collaboration}),\ }\bibfield  {title} {\bibinfo {title} {Broadband quantum enhancement of the ligo detectors with frequency-dependent squeezing},\ }\href {https://doi.org/10.1103/PhysRevX.13.041021} {\bibfield  {journal} {\bibinfo  {journal} {Phys. Rev. X}\ }\textbf {\bibinfo {volume} {13}},\ \bibinfo {pages} {041021} (\bibinfo {year} {2023})}\BibitemShut {NoStop}%
\bibitem [{\citenamefont {Whittle}\ \emph {et~al.}(2021)\citenamefont {Whittle} \emph {et~al.}}]{Whittle2021}%
  \BibitemOpen
  \bibfield  {author} {\bibinfo {author} {\bibfnamefont {C.}~\bibnamefont {Whittle}} \emph {et~al.},\ }\bibfield  {title} {\bibinfo {title} {Approaching the motional ground state of a 10-kg object},\ }\href {https://doi.org/10.1126/science.abh2634} {\bibfield  {journal} {\bibinfo  {journal} {Science}\ }\textbf {\bibinfo {volume} {372}},\ \bibinfo {pages} {1333} (\bibinfo {year} {2021})},\ \Eprint {https://arxiv.org/abs/https://www.science.org/doi/pdf/10.1126/science.abh2634} {https://www.science.org/doi/pdf/10.1126/science.abh2634} \BibitemShut {NoStop}%
\bibitem [{\citenamefont {Tebbenjohanns}\ \emph {et~al.}(2019)\citenamefont {Tebbenjohanns}, \citenamefont {Frimmer},\ and\ \citenamefont {Novotny}}]{Tebbenjohanns_Frimmer_Novotny_2019}%
  \BibitemOpen
  \bibfield  {author} {\bibinfo {author} {\bibfnamefont {F.}~\bibnamefont {Tebbenjohanns}}, \bibinfo {author} {\bibfnamefont {M.}~\bibnamefont {Frimmer}},\ and\ \bibinfo {author} {\bibfnamefont {L.}~\bibnamefont {Novotny}},\ }\bibfield  {title} {\bibinfo {title} {Optimal position detection of a dipolar scatterer in a focused field},\ }\href {https://doi.org/10.1103/PhysRevA.100.043821} {\bibfield  {journal} {\bibinfo  {journal} {Physical Review A}\ }\textbf {\bibinfo {volume} {100}},\ \bibinfo {pages} {043821} (\bibinfo {year} {2019})}\BibitemShut {NoStop}%
\bibitem [{\citenamefont {Gonzalez-Ballestero}\ \emph {et~al.}(2021)\citenamefont {Gonzalez-Ballestero}, \citenamefont {Aspelmeyer}, \citenamefont {Novotny}, \citenamefont {Quidant},\ and\ \citenamefont {Romero-Isart}}]{Gonzalez-Ballestero2021}%
  \BibitemOpen
  \bibfield  {author} {\bibinfo {author} {\bibfnamefont {C.}~\bibnamefont {Gonzalez-Ballestero}}, \bibinfo {author} {\bibfnamefont {M.}~\bibnamefont {Aspelmeyer}}, \bibinfo {author} {\bibfnamefont {L.}~\bibnamefont {Novotny}}, \bibinfo {author} {\bibfnamefont {R.}~\bibnamefont {Quidant}},\ and\ \bibinfo {author} {\bibfnamefont {O.}~\bibnamefont {Romero-Isart}},\ }\bibfield  {title} {\bibinfo {title} {Levitodynamics: Levitation and control of microscopic objects in vacuum},\ }\href {https://doi.org/10.1126/science.abg3027} {\bibfield  {journal} {\bibinfo  {journal} {Science}\ }\textbf {\bibinfo {volume} {374}},\ \bibinfo {pages} {eabg3027} (\bibinfo {year} {2021})},\ \Eprint {https://arxiv.org/abs/https://www.science.org/doi/pdf/10.1126/science.abg3027} {https://www.science.org/doi/pdf/10.1126/science.abg3027} \BibitemShut {NoStop}%
\bibitem [{\citenamefont {Hao}\ and\ \citenamefont {Purdy}(2024)}]{Hao_Purdy_2024}%
  \BibitemOpen
  \bibfield  {author} {\bibinfo {author} {\bibfnamefont {S.}~\bibnamefont {Hao}}\ and\ \bibinfo {author} {\bibfnamefont {T.~P.}\ \bibnamefont {Purdy}},\ }\bibfield  {title} {\bibinfo {title} {Back action evasion in optical lever detection},\ }\href {https://doi.org/10.1364/OPTICA.500036} {\bibfield  {journal} {\bibinfo  {journal} {Optica}\ }\textbf {\bibinfo {volume} {11}},\ \bibinfo {pages} {10–17} (\bibinfo {year} {2024})}\BibitemShut {NoStop}%
\bibitem [{\citenamefont {Pluchar}\ \emph {et~al.}(2025)\citenamefont {Pluchar}, \citenamefont {He}, \citenamefont {Manley}, \citenamefont {Deshler}, \citenamefont {Guha},\ and\ \citenamefont {Wilson}}]{Pluchar_He_Manley_Deshler_Guha_Wilson_2025}%
  \BibitemOpen
  \bibfield  {author} {\bibinfo {author} {\bibfnamefont {C.~M.}\ \bibnamefont {Pluchar}}, \bibinfo {author} {\bibfnamefont {W.}~\bibnamefont {He}}, \bibinfo {author} {\bibfnamefont {J.}~\bibnamefont {Manley}}, \bibinfo {author} {\bibfnamefont {N.}~\bibnamefont {Deshler}}, \bibinfo {author} {\bibfnamefont {S.}~\bibnamefont {Guha}},\ and\ \bibinfo {author} {\bibfnamefont {D.~J.}\ \bibnamefont {Wilson}},\ }\bibfield  {title} {\bibinfo {title} {Imaging-based quantum optomechanics},\ }\href {https://doi.org/10.1103/64xv-3fyx} {\bibfield  {journal} {\bibinfo  {journal} {Physical Review Letters}\ }\textbf {\bibinfo {volume} {135}},\ \bibinfo {pages} {023601} (\bibinfo {year} {2025})}\BibitemShut {NoStop}%
\bibitem [{\citenamefont {Pratt}\ \emph {et~al.}(2023)\citenamefont {Pratt}, \citenamefont {Agrawal}, \citenamefont {Condos}, \citenamefont {Pluchar}, \citenamefont {Schlamminger},\ and\ \citenamefont {Wilson}}]{Pratt_Agrawal_Condos_Pluchar_Schlamminger_Wilson_2023}%
  \BibitemOpen
  \bibfield  {author} {\bibinfo {author} {\bibfnamefont {J.~R.}\ \bibnamefont {Pratt}}, \bibinfo {author} {\bibfnamefont {A.~R.}\ \bibnamefont {Agrawal}}, \bibinfo {author} {\bibfnamefont {C.~A.}\ \bibnamefont {Condos}}, \bibinfo {author} {\bibfnamefont {C.~M.}\ \bibnamefont {Pluchar}}, \bibinfo {author} {\bibfnamefont {S.}~\bibnamefont {Schlamminger}},\ and\ \bibinfo {author} {\bibfnamefont {D.~J.}\ \bibnamefont {Wilson}},\ }\bibfield  {title} {\bibinfo {title} {Nanoscale torsional dissipation dilution for quantum experiments and precision measurement},\ }\href {https://doi.org/10.1103/PhysRevX.13.011018} {\bibfield  {journal} {\bibinfo  {journal} {Physical Review X}\ }\textbf {\bibinfo {volume} {13}},\ \bibinfo {pages} {011018} (\bibinfo {year} {2023})}\BibitemShut {NoStop}%
\bibitem [{\citenamefont {Tay}\ \emph {et~al.}(2008)\citenamefont {Tay}, \citenamefont {Thibierge}, \citenamefont {Fournier}, \citenamefont {Fretigny}, \citenamefont {Lequeux}, \citenamefont {Monteux}, \citenamefont {Roger},\ and\ \citenamefont {Talini}}]{Tay2008}%
  \BibitemOpen
  \bibfield  {author} {\bibinfo {author} {\bibfnamefont {A.}~\bibnamefont {Tay}}, \bibinfo {author} {\bibfnamefont {C.}~\bibnamefont {Thibierge}}, \bibinfo {author} {\bibfnamefont {D.}~\bibnamefont {Fournier}}, \bibinfo {author} {\bibfnamefont {C.}~\bibnamefont {Fretigny}}, \bibinfo {author} {\bibfnamefont {F.}~\bibnamefont {Lequeux}}, \bibinfo {author} {\bibfnamefont {C.}~\bibnamefont {Monteux}}, \bibinfo {author} {\bibfnamefont {J.~P.}\ \bibnamefont {Roger}},\ and\ \bibinfo {author} {\bibfnamefont {L.}~\bibnamefont {Talini}},\ }\bibfield  {title} {\bibinfo {title} {Probing thermal waves on the free surface of various media: Surface fluctuation specular reflection spectroscopy},\ }\href {https://doi.org/10.1063/1.3002424} {\bibfield  {journal} {\bibinfo  {journal} {Review of Scientific Instruments}\ }\textbf {\bibinfo {volume} {79}},\ \bibinfo {pages} {103107} (\bibinfo {year} {2008})}\BibitemShut {NoStop}%
\bibitem [{\citenamefont {Putman}\ \emph {et~al.}(1992)\citenamefont {Putman}, \citenamefont {de~Grooth}, \citenamefont {van Hulst},\ and\ \citenamefont {Greve}}]{Putman_de_Grooth_van_Hulst_Greve_1992}%
  \BibitemOpen
  \bibfield  {author} {\bibinfo {author} {\bibfnamefont {C.~A.~J.}\ \bibnamefont {Putman}}, \bibinfo {author} {\bibfnamefont {B.~G.}\ \bibnamefont {de~Grooth}}, \bibinfo {author} {\bibfnamefont {N.~F.}\ \bibnamefont {van Hulst}},\ and\ \bibinfo {author} {\bibfnamefont {J.}~\bibnamefont {Greve}},\ }\bibfield  {title} {\bibinfo {title} {A theoretical comparison between interferometric and optical beam deflection technique for the measurement of cantilever displacement in afm},\ }\href {https://doi.org/10.1016/0304-3991(92)90474-X} {\bibfield  {journal} {\bibinfo  {journal} {Ultramicroscopy}\ }\textbf {\bibinfo {volume} {42–44}},\ \bibinfo {pages} {1509–1513} (\bibinfo {year} {1992})}\BibitemShut {NoStop}%
\bibitem [{\citenamefont {Aspelmeyer}\ \emph {et~al.}(2014)\citenamefont {Aspelmeyer}, \citenamefont {Kippenberg},\ and\ \citenamefont {Marquardt}}]{Aspelmeyer_Kippenberg_Marquardt_2014}%
  \BibitemOpen
  \bibfield  {author} {\bibinfo {author} {\bibfnamefont {M.}~\bibnamefont {Aspelmeyer}}, \bibinfo {author} {\bibfnamefont {T.~J.}\ \bibnamefont {Kippenberg}},\ and\ \bibinfo {author} {\bibfnamefont {F.}~\bibnamefont {Marquardt}},\ }\bibfield  {title} {\bibinfo {title} {Cavity optomechanics},\ }\href {https://doi.org/10.1103/RevModPhys.86.1391} {\bibfield  {journal} {\bibinfo  {journal} {Reviews of Modern Physics}\ }\textbf {\bibinfo {volume} {86}},\ \bibinfo {pages} {1391–1452} (\bibinfo {year} {2014})}\BibitemShut {NoStop}%
\bibitem [{\citenamefont {Clerk}\ \emph {et~al.}(2010)\citenamefont {Clerk}, \citenamefont {Devoret}, \citenamefont {Girvin}, \citenamefont {Marquardt},\ and\ \citenamefont {Schoelkopf}}]{Clerk_Devoret_Girvin_Marquardt_Schoelkopf_2010}%
  \BibitemOpen
  \bibfield  {author} {\bibinfo {author} {\bibfnamefont {A.~A.}\ \bibnamefont {Clerk}}, \bibinfo {author} {\bibfnamefont {M.~H.}\ \bibnamefont {Devoret}}, \bibinfo {author} {\bibfnamefont {S.~M.}\ \bibnamefont {Girvin}}, \bibinfo {author} {\bibfnamefont {F.}~\bibnamefont {Marquardt}},\ and\ \bibinfo {author} {\bibfnamefont {R.~J.}\ \bibnamefont {Schoelkopf}},\ }\bibfield  {title} {\bibinfo {title} {Introduction to quantum noise, measurement, and amplification},\ }\href {https://doi.org/10.1103/RevModPhys.82.1155} {\bibfield  {journal} {\bibinfo  {journal} {Reviews of Modern Physics}\ }\textbf {\bibinfo {volume} {82}},\ \bibinfo {pages} {1155–1208} (\bibinfo {year} {2010})}\BibitemShut {NoStop}%
\bibitem [{\citenamefont {Maurer}\ \emph {et~al.}(2023)\citenamefont {Maurer}, \citenamefont {Gonzalez-Ballestero},\ and\ \citenamefont {Romero-Isart}}]{Maurer_Gonzalez-Ballestero_Romero-Isart_2023}%
  \BibitemOpen
  \bibfield  {author} {\bibinfo {author} {\bibfnamefont {P.}~\bibnamefont {Maurer}}, \bibinfo {author} {\bibfnamefont {C.}~\bibnamefont {Gonzalez-Ballestero}},\ and\ \bibinfo {author} {\bibfnamefont {O.}~\bibnamefont {Romero-Isart}},\ }\bibfield  {title} {\bibinfo {title} {Quantum theory of light interaction with a lorenz-mie particle: Optical detection and three-dimensional ground-state cooling},\ }\href {https://doi.org/10.1103/PhysRevA.108.033714} {\bibfield  {journal} {\bibinfo  {journal} {Physical Review A}\ }\textbf {\bibinfo {volume} {108}},\ \bibinfo {pages} {033714} (\bibinfo {year} {2023})}\BibitemShut {NoStop}%
\bibitem [{\citenamefont {H{\"u}pfl}\ \emph {et~al.}(2024)\citenamefont {H{\"u}pfl}, \citenamefont {Russo}, \citenamefont {Rachbauer}, \citenamefont {Bouchet}, \citenamefont {Lu}, \citenamefont {Kuhl},\ and\ \citenamefont {Rotter}}]{Hupfl_Russo_Rachbauer_Bouchet_Lu_Kuhl_Rotter_2024}%
  \BibitemOpen
  \bibfield  {author} {\bibinfo {author} {\bibfnamefont {J.}~\bibnamefont {H{\"u}pfl}}, \bibinfo {author} {\bibfnamefont {F.}~\bibnamefont {Russo}}, \bibinfo {author} {\bibfnamefont {L.~M.}\ \bibnamefont {Rachbauer}}, \bibinfo {author} {\bibfnamefont {D.}~\bibnamefont {Bouchet}}, \bibinfo {author} {\bibfnamefont {J.}~\bibnamefont {Lu}}, \bibinfo {author} {\bibfnamefont {U.}~\bibnamefont {Kuhl}},\ and\ \bibinfo {author} {\bibfnamefont {S.}~\bibnamefont {Rotter}},\ }\bibfield  {title} {\bibinfo {title} {Continuity equation for the flow of fisher information in wave scattering},\ }\href {https://doi.org/10.1038/s41567-024-02519-8} {\bibfield  {journal} {\bibinfo  {journal} {Nature Physics}\ }\textbf {\bibinfo {volume} {20}},\ \bibinfo {pages} {1294–1299} (\bibinfo {year} {2024})}\BibitemShut {NoStop}%
\bibitem [{\citenamefont {Dania}\ \emph {et~al.}(2022)\citenamefont {Dania}, \citenamefont {Heidegger}, \citenamefont {Bykov}, \citenamefont {Cerchiari}, \citenamefont {Araneda},\ and\ \citenamefont {Northup}}]{Dania_Heidegger_Bykov_Cerchiari_Araneda_Northup_2022}%
  \BibitemOpen
  \bibfield  {author} {\bibinfo {author} {\bibfnamefont {L.}~\bibnamefont {Dania}}, \bibinfo {author} {\bibfnamefont {K.}~\bibnamefont {Heidegger}}, \bibinfo {author} {\bibfnamefont {D.~S.}\ \bibnamefont {Bykov}}, \bibinfo {author} {\bibfnamefont {G.}~\bibnamefont {Cerchiari}}, \bibinfo {author} {\bibfnamefont {G.}~\bibnamefont {Araneda}},\ and\ \bibinfo {author} {\bibfnamefont {T.~E.}\ \bibnamefont {Northup}},\ }\bibfield  {title} {\bibinfo {title} {Position measurement of a levitated nanoparticle via interference with its mirror image},\ }\href {https://doi.org/10.1103/PhysRevLett.129.013601} {\bibfield  {journal} {\bibinfo  {journal} {Physical Review Letters}\ }\textbf {\bibinfo {volume} {129}},\ \bibinfo {pages} {013601} (\bibinfo {year} {2022})}\BibitemShut {NoStop}%
\bibitem [{\citenamefont {Tebbenjohanns}\ \emph {et~al.}(2021)\citenamefont {Tebbenjohanns}, \citenamefont {Mattana}, \citenamefont {Rossi}, \citenamefont {Frimmer},\ and\ \citenamefont {Novotny}}]{Tebbenjohanns2021}%
  \BibitemOpen
  \bibfield  {author} {\bibinfo {author} {\bibfnamefont {F.}~\bibnamefont {Tebbenjohanns}}, \bibinfo {author} {\bibfnamefont {M.~L.}\ \bibnamefont {Mattana}}, \bibinfo {author} {\bibfnamefont {M.}~\bibnamefont {Rossi}}, \bibinfo {author} {\bibfnamefont {M.}~\bibnamefont {Frimmer}},\ and\ \bibinfo {author} {\bibfnamefont {L.}~\bibnamefont {Novotny}},\ }\bibfield  {title} {\bibinfo {title} {Quantum control of a nanoparticle optically levitated in cryogenic free space},\ }\href@noop {} {\bibfield  {journal} {\bibinfo  {journal} {Nature}\ }\textbf {\bibinfo {volume} {595}},\ \bibinfo {pages} {378} (\bibinfo {year} {2021})}\BibitemShut {NoStop}%
\bibitem [{\citenamefont {Magrini}\ \emph {et~al.}(2021)\citenamefont {Magrini}, \citenamefont {Rosenzweig}, \citenamefont {Bach}, \citenamefont {Deutschmann-Olek}, \citenamefont {Hofer}, \citenamefont {Hong}, \citenamefont {Kiesel}, \citenamefont {Kugi},\ and\ \citenamefont {Aspelmeyer}}]{Magrini2021}%
  \BibitemOpen
  \bibfield  {author} {\bibinfo {author} {\bibfnamefont {L.}~\bibnamefont {Magrini}}, \bibinfo {author} {\bibfnamefont {P.}~\bibnamefont {Rosenzweig}}, \bibinfo {author} {\bibfnamefont {C.}~\bibnamefont {Bach}}, \bibinfo {author} {\bibfnamefont {A.}~\bibnamefont {Deutschmann-Olek}}, \bibinfo {author} {\bibfnamefont {S.~G.}\ \bibnamefont {Hofer}}, \bibinfo {author} {\bibfnamefont {S.}~\bibnamefont {Hong}}, \bibinfo {author} {\bibfnamefont {N.}~\bibnamefont {Kiesel}}, \bibinfo {author} {\bibfnamefont {A.}~\bibnamefont {Kugi}},\ and\ \bibinfo {author} {\bibfnamefont {M.}~\bibnamefont {Aspelmeyer}},\ }\bibfield  {title} {\bibinfo {title} {Real-time optimal quantum control of mechanical motion at room temperature},\ }\href@noop {} {\bibfield  {journal} {\bibinfo  {journal} {Nature}\ }\textbf {\bibinfo {volume} {595}},\ \bibinfo {pages} {373} (\bibinfo {year} {2021})}\BibitemShut {NoStop}%
\bibitem [{\citenamefont {Tebbenjohanns}\ \emph {et~al.}(2022)\citenamefont {Tebbenjohanns}, \citenamefont {Militaru}, \citenamefont {Norrman}, \citenamefont {van~der Laan}, \citenamefont {Novotny},\ and\ \citenamefont {Frimmer}}]{Tebbenjohanns2022}%
  \BibitemOpen
  \bibfield  {author} {\bibinfo {author} {\bibfnamefont {F.}~\bibnamefont {Tebbenjohanns}}, \bibinfo {author} {\bibfnamefont {A.}~\bibnamefont {Militaru}}, \bibinfo {author} {\bibfnamefont {A.}~\bibnamefont {Norrman}}, \bibinfo {author} {\bibfnamefont {F.}~\bibnamefont {van~der Laan}}, \bibinfo {author} {\bibfnamefont {L.}~\bibnamefont {Novotny}},\ and\ \bibinfo {author} {\bibfnamefont {M.}~\bibnamefont {Frimmer}},\ }\bibfield  {title} {\bibinfo {title} {Optimal orientation detection of an anisotropic dipolar scatterer},\ }\href {https://doi.org/10.1103/PhysRevA.105.053504} {\bibfield  {journal} {\bibinfo  {journal} {Phys. Rev. A}\ }\textbf {\bibinfo {volume} {105}},\ \bibinfo {pages} {053504} (\bibinfo {year} {2022})}\BibitemShut {NoStop}%
\bibitem [{\citenamefont {Laing}\ \emph {et~al.}(2024)\citenamefont {Laing}, \citenamefont {Klomp}, \citenamefont {Winstone}, \citenamefont {Grinin}, \citenamefont {Dana}, \citenamefont {Wang}, \citenamefont {Widyatmodjo}, \citenamefont {Bateman},\ and\ \citenamefont {Geraci}}]{Laing_Klomp_Winstone_Grinin_Dana_Wang_Widyatmodjo_Bateman_Geraci_2024}%
  \BibitemOpen
  \bibfield  {author} {\bibinfo {author} {\bibfnamefont {S.}~\bibnamefont {Laing}}, \bibinfo {author} {\bibfnamefont {S.}~\bibnamefont {Klomp}}, \bibinfo {author} {\bibfnamefont {G.}~\bibnamefont {Winstone}}, \bibinfo {author} {\bibfnamefont {A.}~\bibnamefont {Grinin}}, \bibinfo {author} {\bibfnamefont {A.}~\bibnamefont {Dana}}, \bibinfo {author} {\bibfnamefont {Z.}~\bibnamefont {Wang}}, \bibinfo {author} {\bibfnamefont {K.~S.}\ \bibnamefont {Widyatmodjo}}, \bibinfo {author} {\bibfnamefont {J.}~\bibnamefont {Bateman}},\ and\ \bibinfo {author} {\bibfnamefont {A.~A.}\ \bibnamefont {Geraci}},\ }\bibfield  {title} {\bibinfo {title} {Optimal displacement detection of arbitrarily-shaped levitated dielectric objects using optical radiation}\ }\href {https://doi.org/10.48550/arXiv.2409.00782} {10.48550/arXiv.2409.00782} (\bibinfo {year} {2024}),\ \bibinfo {note} {arXiv:2409.00782 [physics]}\BibitemShut {NoStop}%
\bibitem [{\citenamefont {Gao}\ \emph {et~al.}(2024)\citenamefont {Gao}, \citenamefont {van~der Laan}, \citenamefont {Zielińska}, \citenamefont {Militaru}, \citenamefont {Novotny},\ and\ \citenamefont {Frimmer}}]{Gao_van_der_Laan_Zielińska_Militaru_Novotny_Frimmer_2024}%
  \BibitemOpen
  \bibfield  {author} {\bibinfo {author} {\bibfnamefont {J.}~\bibnamefont {Gao}}, \bibinfo {author} {\bibfnamefont {F.}~\bibnamefont {van~der Laan}}, \bibinfo {author} {\bibfnamefont {J.~A.}\ \bibnamefont {Zielińska}}, \bibinfo {author} {\bibfnamefont {A.}~\bibnamefont {Militaru}}, \bibinfo {author} {\bibfnamefont {L.}~\bibnamefont {Novotny}},\ and\ \bibinfo {author} {\bibfnamefont {M.}~\bibnamefont {Frimmer}},\ }\bibfield  {title} {\bibinfo {title} {Feedback cooling a levitated nanoparticle’s libration to below 100 phonons},\ }\href {https://doi.org/10.1103/PhysRevResearch.6.033009} {\bibfield  {journal} {\bibinfo  {journal} {Physical Review Research}\ }\textbf {\bibinfo {volume} {6}},\ \bibinfo {pages} {033009} (\bibinfo {year} {2024})}\BibitemShut {NoStop}%
\bibitem [{\citenamefont {Bang}\ \emph {et~al.}(2020)\citenamefont {Bang}, \citenamefont {Seberson}, \citenamefont {Ju}, \citenamefont {Ahn}, \citenamefont {Xu}, \citenamefont {Gao}, \citenamefont {Robicheaux},\ and\ \citenamefont {Li}}]{Bang2020}%
  \BibitemOpen
  \bibfield  {author} {\bibinfo {author} {\bibfnamefont {J.}~\bibnamefont {Bang}}, \bibinfo {author} {\bibfnamefont {T.}~\bibnamefont {Seberson}}, \bibinfo {author} {\bibfnamefont {P.}~\bibnamefont {Ju}}, \bibinfo {author} {\bibfnamefont {J.}~\bibnamefont {Ahn}}, \bibinfo {author} {\bibfnamefont {Z.}~\bibnamefont {Xu}}, \bibinfo {author} {\bibfnamefont {X.}~\bibnamefont {Gao}}, \bibinfo {author} {\bibfnamefont {F.}~\bibnamefont {Robicheaux}},\ and\ \bibinfo {author} {\bibfnamefont {T.}~\bibnamefont {Li}},\ }\bibfield  {title} {\bibinfo {title} {Five-dimensional cooling and nonlinear dynamics of an optically levitated nanodumbbell},\ }\href {https://doi.org/10.1103/PhysRevResearch.2.043054} {\bibfield  {journal} {\bibinfo  {journal} {Phys. Rev. Res.}\ }\textbf {\bibinfo {volume} {2}},\ \bibinfo {pages} {043054} (\bibinfo {year} {2020})}\BibitemShut {NoStop}%
\bibitem [{\citenamefont {Rossi}\ \emph {et~al.}(2018)\citenamefont {Rossi}, \citenamefont {Mason}, \citenamefont {Chen}, \citenamefont {Tsaturyan},\ and\ \citenamefont {Schliesser}}]{Rossi2018}%
  \BibitemOpen
  \bibfield  {author} {\bibinfo {author} {\bibfnamefont {M.}~\bibnamefont {Rossi}}, \bibinfo {author} {\bibfnamefont {D.}~\bibnamefont {Mason}}, \bibinfo {author} {\bibfnamefont {J.}~\bibnamefont {Chen}}, \bibinfo {author} {\bibfnamefont {Y.}~\bibnamefont {Tsaturyan}},\ and\ \bibinfo {author} {\bibfnamefont {A.}~\bibnamefont {Schliesser}},\ }\bibfield  {title} {\bibinfo {title} {Measurement-based quantum control of mechanical motion},\ }\href@noop {} {\bibfield  {journal} {\bibinfo  {journal} {Nature}\ }\textbf {\bibinfo {volume} {563}},\ \bibinfo {pages} {53} (\bibinfo {year} {2018})}\BibitemShut {NoStop}%
\bibitem [{\citenamefont {Shin}\ \emph {et~al.}(2025)\citenamefont {Shin}, \citenamefont {Hayward}, \citenamefont {Fife}, \citenamefont {Menon},\ and\ \citenamefont {Sudhir}}]{Shin2025}%
  \BibitemOpen
  \bibfield  {author} {\bibinfo {author} {\bibfnamefont {D.-C.}\ \bibnamefont {Shin}}, \bibinfo {author} {\bibfnamefont {T.~M.}\ \bibnamefont {Hayward}}, \bibinfo {author} {\bibfnamefont {D.}~\bibnamefont {Fife}}, \bibinfo {author} {\bibfnamefont {R.}~\bibnamefont {Menon}},\ and\ \bibinfo {author} {\bibfnamefont {V.}~\bibnamefont {Sudhir}},\ }\bibfield  {title} {\bibinfo {title} {Active laser cooling of a centimeter-scale torsional oscillator},\ }\href {https://doi.org/10.1364/OPTICA.548098} {\bibfield  {journal} {\bibinfo  {journal} {Optica}\ }\textbf {\bibinfo {volume} {12}},\ \bibinfo {pages} {473} (\bibinfo {year} {2025})}\BibitemShut {NoStop}%
\bibitem [{\citenamefont {Bouchet}\ \emph {et~al.}(2021)\citenamefont {Bouchet}, \citenamefont {Rotter},\ and\ \citenamefont {Mosk}}]{Bouchet_Rotter_Mosk_2021}%
  \BibitemOpen
  \bibfield  {author} {\bibinfo {author} {\bibfnamefont {D.}~\bibnamefont {Bouchet}}, \bibinfo {author} {\bibfnamefont {S.}~\bibnamefont {Rotter}},\ and\ \bibinfo {author} {\bibfnamefont {A.~P.}\ \bibnamefont {Mosk}},\ }\bibfield  {title} {\bibinfo {title} {Maximum information states for coherent scattering measurements},\ }\href {https://doi.org/10.1038/s41567-020-01137-4} {\bibfield  {journal} {\bibinfo  {journal} {Nature Physics}\ }\textbf {\bibinfo {volume} {17}},\ \bibinfo {pages} {564–568} (\bibinfo {year} {2021})}\BibitemShut {NoStop}%
\bibitem [{sup()}]{supp}%
  \BibitemOpen
  \href@noop {} {}\bibinfo {note} {See Supplemental Material [] for additional theoretical and experimental details}\BibitemShut {NoStop}%
\bibitem [{\citenamefont {Enomoto}\ \emph {et~al.}(2016)\citenamefont {Enomoto}, \citenamefont {Nagano},\ and\ \citenamefont {Kawamura}}]{Enomoto2016}%
  \BibitemOpen
  \bibfield  {author} {\bibinfo {author} {\bibfnamefont {Y.}~\bibnamefont {Enomoto}}, \bibinfo {author} {\bibfnamefont {K.}~\bibnamefont {Nagano}},\ and\ \bibinfo {author} {\bibfnamefont {S.}~\bibnamefont {Kawamura}},\ }\bibfield  {title} {\bibinfo {title} {Standard quantum limit of angular motion of a suspended mirror and homodyne detection of a ponderomotively squeezed vacuum field},\ }\href {https://doi.org/10.1103/PhysRevA.94.012115} {\bibfield  {journal} {\bibinfo  {journal} {Phys. Rev. A}\ }\textbf {\bibinfo {volume} {94}},\ \bibinfo {pages} {012115} (\bibinfo {year} {2016})}\BibitemShut {NoStop}%
\bibitem [{\citenamefont {Pinard}\ \emph {et~al.}(1999)\citenamefont {Pinard}, \citenamefont {Hadjar},\ and\ \citenamefont {Heidmann}}]{Pinard1999}%
  \BibitemOpen
  \bibfield  {author} {\bibinfo {author} {\bibfnamefont {M.}~\bibnamefont {Pinard}}, \bibinfo {author} {\bibfnamefont {Y.}~\bibnamefont {Hadjar}},\ and\ \bibinfo {author} {\bibfnamefont {A.}~\bibnamefont {Heidmann}},\ }\bibfield  {title} {\bibinfo {title} {Effective mass in quantum effects of radiation pressure},\ }\href@noop {} {\bibfield  {journal} {\bibinfo  {journal} {The European Physical Journal D-Atomic, Molecular, Optical and Plasma Physics}\ }\textbf {\bibinfo {volume} {7}},\ \bibinfo {pages} {107} (\bibinfo {year} {1999})}\BibitemShut {NoStop}%
\bibitem [{\citenamefont {Knee}\ and\ \citenamefont {Munro}(2015)}]{Knee2015}%
  \BibitemOpen
  \bibfield  {author} {\bibinfo {author} {\bibfnamefont {G.~C.}\ \bibnamefont {Knee}}\ and\ \bibinfo {author} {\bibfnamefont {W.~J.}\ \bibnamefont {Munro}},\ }\bibfield  {title} {\bibinfo {title} {Fisher information versus signal-to-noise ratio for a split detector},\ }\href {https://doi.org/10.1103/PhysRevA.92.012130} {\bibfield  {journal} {\bibinfo  {journal} {Phys. Rev. A}\ }\textbf {\bibinfo {volume} {92}},\ \bibinfo {pages} {012130} (\bibinfo {year} {2015})}\BibitemShut {NoStop}%
\bibitem [{\citenamefont {Walborn}\ \emph {et~al.}(2020)\citenamefont {Walborn}, \citenamefont {Aguilar}, \citenamefont {Saldanha}, \citenamefont {Davidovich},\ and\ \citenamefont {Filho}}]{Walborn_Aguilar_Saldanha_Davidovich_Filho_2020}%
  \BibitemOpen
  \bibfield  {author} {\bibinfo {author} {\bibfnamefont {S.~P.}\ \bibnamefont {Walborn}}, \bibinfo {author} {\bibfnamefont {G.~H.}\ \bibnamefont {Aguilar}}, \bibinfo {author} {\bibfnamefont {P.~L.}\ \bibnamefont {Saldanha}}, \bibinfo {author} {\bibfnamefont {L.}~\bibnamefont {Davidovich}},\ and\ \bibinfo {author} {\bibfnamefont {R.~L. d.~M.}\ \bibnamefont {Filho}},\ }\bibfield  {title} {\bibinfo {title} {Interferometric sensing of the tilt angle of a gaussian beam},\ }\href {https://doi.org/10.1103/PhysRevResearch.2.033191} {\bibfield  {journal} {\bibinfo  {journal} {Physical Review Research}\ }\textbf {\bibinfo {volume} {2}},\ \bibinfo {pages} {033191} (\bibinfo {year} {2020})}\BibitemShut {NoStop}%
\bibitem [{\citenamefont {Hsu}\ \emph {et~al.}(2004)\citenamefont {Hsu}, \citenamefont {Delaubert}, \citenamefont {Lam},\ and\ \citenamefont {Bowen}}]{Hsu_Delaubert_Lam_Bowen_2004}%
  \BibitemOpen
  \bibfield  {author} {\bibinfo {author} {\bibfnamefont {M.~T.~L.}\ \bibnamefont {Hsu}}, \bibinfo {author} {\bibfnamefont {V.}~\bibnamefont {Delaubert}}, \bibinfo {author} {\bibfnamefont {P.~K.}\ \bibnamefont {Lam}},\ and\ \bibinfo {author} {\bibfnamefont {W.~P.}\ \bibnamefont {Bowen}},\ }\bibfield  {title} {\bibinfo {title} {Optimal optical measurement of small displacements},\ }\href {https://doi.org/10.1088/1464-4266/6/12/003} {\bibfield  {journal} {\bibinfo  {journal} {Journal of Optics B: Quantum and Semiclassical Optics}\ }\textbf {\bibinfo {volume} {6}},\ \bibinfo {pages} {495} (\bibinfo {year} {2004})}\BibitemShut {NoStop}%
\bibitem [{\citenamefont {Fradgley}\ \emph {et~al.}(2022)\citenamefont {Fradgley}, \citenamefont {French}, \citenamefont {Rushton}, \citenamefont {Dieudonné}, \citenamefont {Harrison}, \citenamefont {Beckey}, \citenamefont {Miao}, \citenamefont {Gill}, \citenamefont {Petrov},\ and\ \citenamefont {Boyer}}]{Fradgley_French_Rushton_Dieudonné_Harrison_Beckey_Miao_Gill_Petrov_Boyer_2022}%
  \BibitemOpen
  \bibfield  {author} {\bibinfo {author} {\bibfnamefont {E.}~\bibnamefont {Fradgley}}, \bibinfo {author} {\bibfnamefont {C.}~\bibnamefont {French}}, \bibinfo {author} {\bibfnamefont {L.}~\bibnamefont {Rushton}}, \bibinfo {author} {\bibfnamefont {Y.}~\bibnamefont {Dieudonné}}, \bibinfo {author} {\bibfnamefont {L.}~\bibnamefont {Harrison}}, \bibinfo {author} {\bibfnamefont {J.~L.}\ \bibnamefont {Beckey}}, \bibinfo {author} {\bibfnamefont {H.}~\bibnamefont {Miao}}, \bibinfo {author} {\bibfnamefont {C.}~\bibnamefont {Gill}}, \bibinfo {author} {\bibfnamefont {P.~G.}\ \bibnamefont {Petrov}},\ and\ \bibinfo {author} {\bibfnamefont {V.}~\bibnamefont {Boyer}},\ }\bibfield  {title} {\bibinfo {title} {Quantum limits of position-sensitive photodiodes},\ }\href {https://doi.org/10.1364/OE.471673} {\bibfield  {journal} {\bibinfo  {journal} {Optics Express}\ }\textbf {\bibinfo {volume} {30}},\ \bibinfo {pages} {39374–39381} (\bibinfo {year} {2022})}\BibitemShut {NoStop}%
\bibitem [{\citenamefont {Dinter}\ \emph {et~al.}(2024)\citenamefont {Dinter}, \citenamefont {Roberts}, \citenamefont {Volz}, \citenamefont {Schmidt},\ and\ \citenamefont {Laplane}}]{Dinter_Roberts_Volz_Schmidt_Laplane_2024}%
  \BibitemOpen
  \bibfield  {author} {\bibinfo {author} {\bibfnamefont {T.}~\bibnamefont {Dinter}}, \bibinfo {author} {\bibfnamefont {R.}~\bibnamefont {Roberts}}, \bibinfo {author} {\bibfnamefont {T.}~\bibnamefont {Volz}}, \bibinfo {author} {\bibfnamefont {M.~K.}\ \bibnamefont {Schmidt}},\ and\ \bibinfo {author} {\bibfnamefont {C.}~\bibnamefont {Laplane}},\ }\bibfield  {title} {\bibinfo {title} {Three-dimensional and selective displacement sensing of a levitated nanoparticle via spatial mode decomposition}\ }\href {https://doi.org/10.48550/arXiv.2409.08827} {10.48550/arXiv.2409.08827} (\bibinfo {year} {2024}),\ \bibinfo {note} {arXiv:2409.08827 [physics]}\BibitemShut {NoStop}%
\bibitem [{\citenamefont {Choi}\ \emph {et~al.}(2024)\citenamefont {Choi}, \citenamefont {Pluchar}, \citenamefont {He}, \citenamefont {Guha},\ and\ \citenamefont {Wilson}}]{Choi_Pluchar_He_Guha_Wilson_2024}%
  \BibitemOpen
  \bibfield  {author} {\bibinfo {author} {\bibfnamefont {M.}~\bibnamefont {Choi}}, \bibinfo {author} {\bibfnamefont {C.}~\bibnamefont {Pluchar}}, \bibinfo {author} {\bibfnamefont {W.}~\bibnamefont {He}}, \bibinfo {author} {\bibfnamefont {S.}~\bibnamefont {Guha}},\ and\ \bibinfo {author} {\bibfnamefont {D.}~\bibnamefont {Wilson}},\ }\bibfield  {title} {\bibinfo {title} {Quantum limited imaging of a nanomechanical resonator with a spatial mode sorter}\ }\href {https://doi.org/10.48550/arXiv.2411.04980} {10.48550/arXiv.2411.04980} (\bibinfo {year} {2024}),\ \bibinfo {note} {arXiv:2411.04980 [quant-ph]}\BibitemShut {NoStop}%
\bibitem [{\citenamefont {Chapalo}\ \emph {et~al.}(2024)\citenamefont {Chapalo}, \citenamefont {Stylianou}, \citenamefont {M{\'e}gret},\ and\ \citenamefont {Theodosiou}}]{Chapalo2024}%
  \BibitemOpen
  \bibfield  {author} {\bibinfo {author} {\bibfnamefont {I.}~\bibnamefont {Chapalo}}, \bibinfo {author} {\bibfnamefont {A.}~\bibnamefont {Stylianou}}, \bibinfo {author} {\bibfnamefont {P.}~\bibnamefont {M{\'e}gret}},\ and\ \bibinfo {author} {\bibfnamefont {A.}~\bibnamefont {Theodosiou}},\ }\bibfield  {title} {\bibinfo {title} {Advances in optical fiber speckle sensing: A comprehensive review},\ }\bibfield  {journal} {\bibinfo  {journal} {Photonics}\ }\textbf {\bibinfo {volume} {11}},\ \href {https://doi.org/10.3390/photonics11040299} {10.3390/photonics11040299} (\bibinfo {year} {2024})\BibitemShut {NoStop}%
\bibitem [{\citenamefont {Lengenfelder}\ \emph {et~al.}(2021)\citenamefont {Lengenfelder}, \citenamefont {Hohmann}, \citenamefont {Sp{\"a}th}, \citenamefont {Scherbaum}, \citenamefont {Wei{\ss}}, \citenamefont {Rupitsch}, \citenamefont {Schmidt}, \citenamefont {Zalevsky},\ and\ \citenamefont {Kl{\"a}mpfl}}]{Lengenfelder2021}%
  \BibitemOpen
  \bibfield  {author} {\bibinfo {author} {\bibfnamefont {B.}~\bibnamefont {Lengenfelder}}, \bibinfo {author} {\bibfnamefont {M.}~\bibnamefont {Hohmann}}, \bibinfo {author} {\bibfnamefont {M.}~\bibnamefont {Sp{\"a}th}}, \bibinfo {author} {\bibfnamefont {D.}~\bibnamefont {Scherbaum}}, \bibinfo {author} {\bibfnamefont {M.}~\bibnamefont {Wei{\ss}}}, \bibinfo {author} {\bibfnamefont {S.~J.}\ \bibnamefont {Rupitsch}}, \bibinfo {author} {\bibfnamefont {M.}~\bibnamefont {Schmidt}}, \bibinfo {author} {\bibfnamefont {Z.}~\bibnamefont {Zalevsky}},\ and\ \bibinfo {author} {\bibfnamefont {F.}~\bibnamefont {Kl{\"a}mpfl}},\ }\bibfield  {title} {\bibinfo {title} {Remote photoacoustic sensing using single speckle analysis by an ultra-fast four quadrant photo-detector},\ }\href@noop {} {\bibfield  {journal} {\bibinfo  {journal} {Sensors}\ }\textbf {\bibinfo {volume} {21}},\ \bibinfo {pages} {2109} (\bibinfo {year} {2021})}\BibitemShut {NoStop}%
\bibitem [{\citenamefont {Hecht}(2017)}]{hecht}%
  \BibitemOpen
  \bibfield  {author} {\bibinfo {author} {\bibfnamefont {E.}~\bibnamefont {Hecht}},\ }\href@noop {} {\emph {\bibinfo {title} {Optics}}},\ \bibinfo {edition} {5th}\ ed.\ (\bibinfo  {publisher} {Addison Wesley},\ \bibinfo {year} {2017})\ Chap.~\bibinfo {chapter} {13}\BibitemShut {NoStop}%
\bibitem [{\citenamefont {Kheifets}\ \emph {et~al.}(2014)\citenamefont {Kheifets}, \citenamefont {Simha}, \citenamefont {Melin}, \citenamefont {Li},\ and\ \citenamefont {Raizen}}]{Kheifets2014}%
  \BibitemOpen
  \bibfield  {author} {\bibinfo {author} {\bibfnamefont {S.}~\bibnamefont {Kheifets}}, \bibinfo {author} {\bibfnamefont {A.}~\bibnamefont {Simha}}, \bibinfo {author} {\bibfnamefont {K.}~\bibnamefont {Melin}}, \bibinfo {author} {\bibfnamefont {T.}~\bibnamefont {Li}},\ and\ \bibinfo {author} {\bibfnamefont {M.~G.}\ \bibnamefont {Raizen}},\ }\bibfield  {title} {\bibinfo {title} {Observation of brownian motion in liquids at short times: Instantaneous velocity and memory loss},\ }\href {https://doi.org/10.1126/science.1248091} {\bibfield  {journal} {\bibinfo  {journal} {Science}\ }\textbf {\bibinfo {volume} {343}},\ \bibinfo {pages} {1493} (\bibinfo {year} {2014})},\ \Eprint {https://arxiv.org/abs/https://www.science.org/doi/pdf/10.1126/science.1248091} {https://www.science.org/doi/pdf/10.1126/science.1248091} \BibitemShut {NoStop}%
\bibitem [{\citenamefont {Militaru}\ \emph {et~al.}(2022)\citenamefont {Militaru}, \citenamefont {Rossi}, \citenamefont {Tebbenjohanns}, \citenamefont {Romero-Isart}, \citenamefont {Frimmer},\ and\ \citenamefont {Novotny}}]{Militaru2022}%
  \BibitemOpen
  \bibfield  {author} {\bibinfo {author} {\bibfnamefont {A.}~\bibnamefont {Militaru}}, \bibinfo {author} {\bibfnamefont {M.}~\bibnamefont {Rossi}}, \bibinfo {author} {\bibfnamefont {F.}~\bibnamefont {Tebbenjohanns}}, \bibinfo {author} {\bibfnamefont {O.}~\bibnamefont {Romero-Isart}}, \bibinfo {author} {\bibfnamefont {M.}~\bibnamefont {Frimmer}},\ and\ \bibinfo {author} {\bibfnamefont {L.}~\bibnamefont {Novotny}},\ }\bibfield  {title} {\bibinfo {title} {Ponderomotive squeezing of light by a levitated nanoparticle in free space},\ }\href {https://doi.org/10.1103/PhysRevLett.129.053602} {\bibfield  {journal} {\bibinfo  {journal} {Phys. Rev. Lett.}\ }\textbf {\bibinfo {volume} {129}},\ \bibinfo {pages} {053602} (\bibinfo {year} {2022})}\BibitemShut {NoStop}%
\bibitem [{\citenamefont {Magrini}\ \emph {et~al.}(2022)\citenamefont {Magrini}, \citenamefont {Camarena-Ch{\'a}vez}, \citenamefont {Bach}, \citenamefont {Johnson},\ and\ \citenamefont {Aspelmeyer}}]{Margrini2022}%
  \BibitemOpen
  \bibfield  {author} {\bibinfo {author} {\bibfnamefont {L.}~\bibnamefont {Magrini}}, \bibinfo {author} {\bibfnamefont {V.~A.}\ \bibnamefont {Camarena-Ch{\'a}vez}}, \bibinfo {author} {\bibfnamefont {C.}~\bibnamefont {Bach}}, \bibinfo {author} {\bibfnamefont {A.}~\bibnamefont {Johnson}},\ and\ \bibinfo {author} {\bibfnamefont {M.}~\bibnamefont {Aspelmeyer}},\ }\bibfield  {title} {\bibinfo {title} {Squeezed light from a levitated nanoparticle at room temperature},\ }\href {https://doi.org/10.1103/PhysRevLett.129.053601} {\bibfield  {journal} {\bibinfo  {journal} {Phys. Rev. Lett.}\ }\textbf {\bibinfo {volume} {129}},\ \bibinfo {pages} {053601} (\bibinfo {year} {2022})}\BibitemShut {NoStop}%
\bibitem [{\citenamefont {Treps}\ \emph {et~al.}(2003)\citenamefont {Treps}, \citenamefont {Grosse}, \citenamefont {Bowen}, \citenamefont {Fabre}, \citenamefont {Bachor},\ and\ \citenamefont {Lam}}]{Treps2003}%
  \BibitemOpen
  \bibfield  {author} {\bibinfo {author} {\bibfnamefont {N.}~\bibnamefont {Treps}}, \bibinfo {author} {\bibfnamefont {N.}~\bibnamefont {Grosse}}, \bibinfo {author} {\bibfnamefont {W.~P.}\ \bibnamefont {Bowen}}, \bibinfo {author} {\bibfnamefont {C.}~\bibnamefont {Fabre}}, \bibinfo {author} {\bibfnamefont {H.-A.}\ \bibnamefont {Bachor}},\ and\ \bibinfo {author} {\bibfnamefont {P.~K.}\ \bibnamefont {Lam}},\ }\bibfield  {title} {\bibinfo {title} {A quantum laser pointer},\ }\href {https://doi.org/10.1126/science.1086489} {\bibfield  {journal} {\bibinfo  {journal} {Science}\ }\textbf {\bibinfo {volume} {301}},\ \bibinfo {pages} {940} (\bibinfo {year} {2003})},\ \Eprint {https://arxiv.org/abs/https://www.science.org/doi/pdf/10.1126/science.1086489} {https://www.science.org/doi/pdf/10.1126/science.1086489} \BibitemShut {NoStop}%
\bibitem [{\citenamefont {Pooser}\ and\ \citenamefont {Lawrie}(2015)}]{Pooser2015}%
  \BibitemOpen
  \bibfield  {author} {\bibinfo {author} {\bibfnamefont {R.~C.}\ \bibnamefont {Pooser}}\ and\ \bibinfo {author} {\bibfnamefont {B.}~\bibnamefont {Lawrie}},\ }\bibfield  {title} {\bibinfo {title} {Ultrasensitive measurement of microcantilever displacement below the shot-noise limit},\ }\href {https://doi.org/10.1364/OPTICA.2.000393} {\bibfield  {journal} {\bibinfo  {journal} {Optica}\ }\textbf {\bibinfo {volume} {2}},\ \bibinfo {pages} {393} (\bibinfo {year} {2015})}\BibitemShut {NoStop}%
\bibitem [{\citenamefont {Habibi}\ \emph {et~al.}(2016)\citenamefont {Habibi}, \citenamefont {Zeuthen}, \citenamefont {Ghanaatshoar},\ and\ \citenamefont {Hammerer}}]{Habibi2016}%
  \BibitemOpen
  \bibfield  {author} {\bibinfo {author} {\bibfnamefont {H.}~\bibnamefont {Habibi}}, \bibinfo {author} {\bibfnamefont {E.}~\bibnamefont {Zeuthen}}, \bibinfo {author} {\bibfnamefont {M.}~\bibnamefont {Ghanaatshoar}},\ and\ \bibinfo {author} {\bibfnamefont {K.}~\bibnamefont {Hammerer}},\ }\bibfield  {title} {\bibinfo {title} {Quantum feedback cooling of a mechanical oscillator using variational measurements: tweaking heisenberg’s microscope},\ }\href@noop {} {\bibfield  {journal} {\bibinfo  {journal} {Journal of Optics}\ }\textbf {\bibinfo {volume} {18}},\ \bibinfo {pages} {084004} (\bibinfo {year} {2016})}\BibitemShut {NoStop}%
\end{thebibliography}%


\begin{thebibliography}{7}%
\makeatletter
\providecommand \@ifxundefined [1]{%
 \@ifx{#1\undefined}
}%
\providecommand \@ifnum [1]{%
 \ifnum #1\expandafter \@firstoftwo
 \else \expandafter \@secondoftwo
 \fi
}%
\providecommand \@ifx [1]{%
 \ifx #1\expandafter \@firstoftwo
 \else \expandafter \@secondoftwo
 \fi
}%
\providecommand \natexlab [1]{#1}%
\providecommand \enquote  [1]{``#1''}%
\providecommand \bibnamefont  [1]{#1}%
\providecommand \bibfnamefont [1]{#1}%
\providecommand \citenamefont [1]{#1}%
\providecommand \href@noop [0]{\@secondoftwo}%
\providecommand \href [0]{\begingroup \@sanitize@url \@href}%
\providecommand \@href[1]{\@@startlink{#1}\@@href}%
\providecommand \@@href[1]{\endgroup#1\@@endlink}%
\providecommand \@sanitize@url [0]{\catcode `\\12\catcode `\$12\catcode `\&12\catcode `\#12\catcode `\^12\catcode `\_12\catcode `\%12\relax}%
\providecommand \@@startlink[1]{}%
\providecommand \@@endlink[0]{}%
\providecommand \url  [0]{\begingroup\@sanitize@url \@url }%
\providecommand \@url [1]{\endgroup\@href {#1}{\urlprefix }}%
\providecommand \urlprefix  [0]{URL }%
\providecommand \Eprint [0]{\href }%
\providecommand \doibase [0]{https://doi.org/}%
\providecommand \selectlanguage [0]{\@gobble}%
\providecommand \bibinfo  [0]{\@secondoftwo}%
\providecommand \bibfield  [0]{\@secondoftwo}%
\providecommand \translation [1]{[#1]}%
\providecommand \BibitemOpen [0]{}%
\providecommand \bibitemStop [0]{}%
\providecommand \bibitemNoStop [0]{.\EOS\space}%
\providecommand \EOS [0]{\spacefactor3000\relax}%
\providecommand \BibitemShut  [1]{\csname bibitem#1\endcsname}%
\let\auto@bib@innerbib\@empty
\bibitem [{\citenamefont {Hecht}(2017)}]{hecht}%
  \BibitemOpen
  \bibfield  {author} {\bibinfo {author} {\bibfnamefont {E.}~\bibnamefont {Hecht}},\ }\href@noop {} {\emph {\bibinfo {title} {Optics}}},\ \bibinfo {edition} {5th}\ ed.\ (\bibinfo  {publisher} {Addison Wesley},\ \bibinfo {year} {2017})\ Chap.~\bibinfo {chapter} {13}\BibitemShut {NoStop}%
\bibitem [{\citenamefont {Fradgley}\ \emph {et~al.}(2022)\citenamefont {Fradgley}, \citenamefont {French}, \citenamefont {Rushton}, \citenamefont {Dieudonné}, \citenamefont {Harrison}, \citenamefont {Beckey}, \citenamefont {Miao}, \citenamefont {Gill}, \citenamefont {Petrov},\ and\ \citenamefont {Boyer}}]{Fradgley_French_Rushton_Dieudonné_Harrison_Beckey_Miao_Gill_Petrov_Boyer_2022}%
  \BibitemOpen
  \bibfield  {author} {\bibinfo {author} {\bibfnamefont {E.}~\bibnamefont {Fradgley}}, \bibinfo {author} {\bibfnamefont {C.}~\bibnamefont {French}}, \bibinfo {author} {\bibfnamefont {L.}~\bibnamefont {Rushton}}, \bibinfo {author} {\bibfnamefont {Y.}~\bibnamefont {Dieudonné}}, \bibinfo {author} {\bibfnamefont {L.}~\bibnamefont {Harrison}}, \bibinfo {author} {\bibfnamefont {J.~L.}\ \bibnamefont {Beckey}}, \bibinfo {author} {\bibfnamefont {H.}~\bibnamefont {Miao}}, \bibinfo {author} {\bibfnamefont {C.}~\bibnamefont {Gill}}, \bibinfo {author} {\bibfnamefont {P.~G.}\ \bibnamefont {Petrov}},\ and\ \bibinfo {author} {\bibfnamefont {V.}~\bibnamefont {Boyer}},\ }\bibfield  {title} {\bibinfo {title} {Quantum limits of position-sensitive photodiodes},\ }\href {https://doi.org/10.1364/OE.471673} {\bibfield  {journal} {\bibinfo  {journal} {Optics Express}\ }\textbf {\bibinfo {volume} {30}},\ \bibinfo {pages} {39374–39381} (\bibinfo {year} {2022})}\BibitemShut {NoStop}%
\bibitem [{\citenamefont {H{\"u}pfl}\ \emph {et~al.}(2024)\citenamefont {H{\"u}pfl}, \citenamefont {Russo}, \citenamefont {Rachbauer}, \citenamefont {Bouchet}, \citenamefont {Lu}, \citenamefont {Kuhl},\ and\ \citenamefont {Rotter}}]{Hupfl_Russo_Rachbauer_Bouchet_Lu_Kuhl_Rotter_2024}%
  \BibitemOpen
  \bibfield  {author} {\bibinfo {author} {\bibfnamefont {J.}~\bibnamefont {H{\"u}pfl}}, \bibinfo {author} {\bibfnamefont {F.}~\bibnamefont {Russo}}, \bibinfo {author} {\bibfnamefont {L.~M.}\ \bibnamefont {Rachbauer}}, \bibinfo {author} {\bibfnamefont {D.}~\bibnamefont {Bouchet}}, \bibinfo {author} {\bibfnamefont {J.}~\bibnamefont {Lu}}, \bibinfo {author} {\bibfnamefont {U.}~\bibnamefont {Kuhl}},\ and\ \bibinfo {author} {\bibfnamefont {S.}~\bibnamefont {Rotter}},\ }\bibfield  {title} {\bibinfo {title} {Continuity equation for the flow of fisher information in wave scattering},\ }\href {https://doi.org/10.1038/s41567-024-02519-8} {\bibfield  {journal} {\bibinfo  {journal} {Nature Physics}\ }\textbf {\bibinfo {volume} {20}},\ \bibinfo {pages} {1294–1299} (\bibinfo {year} {2024})}\BibitemShut {NoStop}%
\bibitem [{\citenamefont {Tebbenjohanns}\ \emph {et~al.}(2019)\citenamefont {Tebbenjohanns}, \citenamefont {Frimmer},\ and\ \citenamefont {Novotny}}]{Tebbenjohanns_Frimmer_Novotny_2019}%
  \BibitemOpen
  \bibfield  {author} {\bibinfo {author} {\bibfnamefont {F.}~\bibnamefont {Tebbenjohanns}}, \bibinfo {author} {\bibfnamefont {M.}~\bibnamefont {Frimmer}},\ and\ \bibinfo {author} {\bibfnamefont {L.}~\bibnamefont {Novotny}},\ }\bibfield  {title} {\bibinfo {title} {Optimal position detection of a dipolar scatterer in a focused field},\ }\href {https://doi.org/10.1103/PhysRevA.100.043821} {\bibfield  {journal} {\bibinfo  {journal} {Physical Review A}\ }\textbf {\bibinfo {volume} {100}},\ \bibinfo {pages} {043821} (\bibinfo {year} {2019})}\BibitemShut {NoStop}%
\bibitem [{\citenamefont {Novotny}(2017)}]{nanooptics}%
  \BibitemOpen
  \bibfield  {author} {\bibinfo {author} {\bibfnamefont {L.}~\bibnamefont {Novotny}},\ }\href@noop {} {\emph {\bibinfo {title} {Principals of Nano-Optics}}},\ \bibinfo {edition} {2nd}\ ed.\ (\bibinfo  {publisher} {Cambridge University Press},\ \bibinfo {year} {2017})\ Chap.~\bibinfo {chapter} {13}\BibitemShut {NoStop}%
\bibitem [{\citenamefont {Bouchet}\ \emph {et~al.}(2021)\citenamefont {Bouchet}, \citenamefont {Rotter},\ and\ \citenamefont {Mosk}}]{Bouchet_Rotter_Mosk_2021}%
  \BibitemOpen
  \bibfield  {author} {\bibinfo {author} {\bibfnamefont {D.}~\bibnamefont {Bouchet}}, \bibinfo {author} {\bibfnamefont {S.}~\bibnamefont {Rotter}},\ and\ \bibinfo {author} {\bibfnamefont {A.~P.}\ \bibnamefont {Mosk}},\ }\bibfield  {title} {\bibinfo {title} {Maximum information states for coherent scattering measurements},\ }\href {https://doi.org/10.1038/s41567-020-01137-4} {\bibfield  {journal} {\bibinfo  {journal} {Nature Physics}\ }\textbf {\bibinfo {volume} {17}},\ \bibinfo {pages} {564–568} (\bibinfo {year} {2021})}\BibitemShut {NoStop}%
\bibitem [{\citenamefont {Pluchar}\ \emph {et~al.}(2025)\citenamefont {Pluchar}, \citenamefont {He}, \citenamefont {Manley}, \citenamefont {Deshler}, \citenamefont {Guha},\ and\ \citenamefont {Wilson}}]{Pluchar_He_Manley_Deshler_Guha_Wilson_2025}%
  \BibitemOpen
  \bibfield  {author} {\bibinfo {author} {\bibfnamefont {C.~M.}\ \bibnamefont {Pluchar}}, \bibinfo {author} {\bibfnamefont {W.}~\bibnamefont {He}}, \bibinfo {author} {\bibfnamefont {J.}~\bibnamefont {Manley}}, \bibinfo {author} {\bibfnamefont {N.}~\bibnamefont {Deshler}}, \bibinfo {author} {\bibfnamefont {S.}~\bibnamefont {Guha}},\ and\ \bibinfo {author} {\bibfnamefont {D.~J.}\ \bibnamefont {Wilson}},\ }\bibfield  {title} {\bibinfo {title} {Imaging-based quantum optomechanics},\ }\href {https://doi.org/10.1103/64xv-3fyx} {\bibfield  {journal} {\bibinfo  {journal} {Physical Review Letters}\ }\textbf {\bibinfo {volume} {135}},\ \bibinfo {pages} {023601} (\bibinfo {year} {2025})}\BibitemShut {NoStop}%
\end{thebibliography}%

\appendix
\section{Theory}\label{AppendixA}
Light with a normalized input mode function $u^{\text{in}}\left(\boldsymbol{r}\right)$ is scattered from a target object in order to extract the value of a parameter $A$, which we take to be a spatial coordinate of the system.  The intensity of the input light is given by $ \left|  \alpha \,  u^{\text{in}}\left(\boldsymbol{r}\right)\right|^2$ , where $\alpha$ is the input coherent state amplitude.  This input beam is scattered from the system, producing an output field mode
\begin{equation}\label{eq:uout}
u^{\text{out}}\left(\boldsymbol{r}\right)\approx u_{0}^{\text{out}}\left(\boldsymbol{r}\right)+kA\, u_{\text{s }}^{\text{out}}\left(\boldsymbol{r}\right),
\end{equation}
where $k$ is the wavenumber of the light, and we assume $kA\ll1$, allowing us to linearize the response.  Here, $u_{0}^{\text{out}}\left(\boldsymbol{r}\right)$  is the stationary output field mode when $A=0$, and $u_{\text{s}}^{\text{out}}\left(\boldsymbol{r}\right)=\frac{1}{k}\frac{du^{\text{out}}}{dA}$ is the (unnormalized) field mode which is populated as $A$ varies.
The light propagates to the detection plane, $D$, where its intensity is:
\[I\left(\boldsymbol{r}\right)= \alpha^2\left(  \left|  u^{\text{out}}_0\left(\boldsymbol{r}\right)\right|^2 +2kA\,\text{Re}\left[ u_{0}^{\text{out}}\left(\boldsymbol{r}\right)u_{\text{s}}^{\text{out}}\left(\boldsymbol{r}\right)^{*}\right] \right)\]
to first order in $A$, where we assume for simplicity that $\alpha$ is real in the detection plane.  Our measurement consists of a weighted sum of photodetectors that we model as a set of domains $D_i$, in the detection plane, representing the extent of each detector and weighting coefficients $f_i$, such that our signal is 
\[ V=\sum_{i}f_i\int_{D_i}I\left(\boldsymbol{r}\right)  da\]
More precisely, $V$ represents the expectation value of signal, and the time variations in $V(t)$ come from the time-varying parameter $A$ and the shot noise intensity fluctuations of the coherent state.  This expression is simplified by defining a weighting function over all of $D$ \[f_w\left(\boldsymbol{r}\right)=\left\{
\begin{aligned}
& f_i ,\boldsymbol{r} \in D_i  \\
& 0, \text{otherwise}
\end{aligned}
\right.\]
 For example, $f_w= \pm 1$ on the opposite halves of a QPD and $f_w=0$ outside the detector.  Then $V=\int_{D}f_w\left(\boldsymbol{r}\right) I\left(\boldsymbol{r}\right)  da$.  To estimate the parameter $A$, we define a sensitivity
 \begin{equation}\label{eq:sensitivity}
 \mathcal{S}=\frac{dV}{dA}=2\alpha^2 k\int_{D}f_w\left(\boldsymbol{r}\right)\text{Re}\left[ u_{0}^{\text{out}}\left(\boldsymbol{r}\right)u_{\text{s}}^{\text{out}}\left(\boldsymbol{r}\right)^{*}\right]da
 \end{equation}
The estimate of $A$ is then $ V/\mathcal{S}$, assuming the expectation value of $V$ is zero when $A=0$, which is true for all the balanced detection schemes considered in this paper. 

To estimate the measurement imprecision, we assume that the noise on each detector is dominated by the shot noise fluctuations on the stationary part, $u_0^{\text{out}}$, of the output field.  We further assume that the shot noise fluctuations are uncorrelated between different detectors.  This assumption will hold for the systems considered in this paper when the detector plane is in the far-field or at an image plane.  However, ponderomotive optomechanical effects can, in general, create spatially dependent optical correlations~\cite{Hao_Purdy_2024, Pluchar_He_Manley_Deshler_Guha_Wilson_2025,Habibi2016}.   We recognize $\bar{N_i}=\int_{D_i} \alpha^2 \left|   u^{\text{out}}_0\left(\boldsymbol{r}\right)\right|^2da$ as the average photon count rate on the $i^{\text{th}}$ detector.  If the detectors have an integration time, $\tau$, that is much shorter than the dynamical time scale of $A$, but long enough that many photons are detected (i.~e.~ $\tau\bar{N_i}\gg 1$ for all $i$), then we can treat the optical noise on each detector as an independent Gaussian distribution with standard deviation $\sqrt{\tau\bar{N_i}}$.  The standard deviation of the signal from each photodetector averaged over the integration time is then given by $\sigma_{V_i}=f_i\sqrt{\tau \bar{N_i}}/\tau$.  We incoherently add the weighted shot noise fluctuations from each detector to compute the variance of the detected signal $ \sigma_V^2=\sum_{i} \sigma_{V_i} ^2=\mathcal{N}/\tau$ where
\begin{equation}\label{eq:noise}
\mathcal{N}=\int_{D}\alpha^2\left|  u^{\text{out}}_0\left(\boldsymbol{r}\right)\right|^2f_w\left(\boldsymbol{r}\right)^2 da.
\end{equation}
This allows us to compute the spectral density of the detected signal $S_V=\mathcal{N}$, for frequencies $\omega<1/\tau$, where it is white, and the spectral density of the imprecision noise of our estimate of $A$, $S_A^{\text{imp}}=\frac{S_V}{\mathcal{S}^2}=\frac{\mathcal{N}}{\mathcal{S}^2}$, which evaluates to
\[S_A^{\text{imp}}=\frac{1}{k^2\alpha^2}\frac{\int_{D}\left|  u^{\text{out}}_0\left(\boldsymbol{r}\right)\right|^2f_w\left(\boldsymbol{r}\right)^2 da}{\left(2\int_{D}f_w\left(\boldsymbol{r}\right)\text{Re}\left[ u_{0}^{\text{out}}\left(\boldsymbol{r}\right)u_{\text{sc}}^{\text{out}}\left(\boldsymbol{r}\right)^{*}\right]da\right)^2}.\]
This results holds under less restrictive conditions than the assumption that many photons land on each detector in the time interval $\tau$.  For a system of many detectors, if the total number of photons detected by all the detectors is large, even if the count rate on an individual detector is small, the central limit theorem is applicable, implying that the noise conforms to the Gaussian distribution given above.  

The measurement efficiency is defined as
\begin{equation}\label{eq:eta}
\eta=\frac{S_A^{\text{ideal}}}{S_A^{\text{imp}}}=S_A^{\text{ideal}}\frac{\mathcal{S}^2}{\mathcal{N}}
\end{equation}
where $S_A^{\text{ideal}}$ is the ideal measurement imprecision that saturates a Heisenberg uncertainty limit, or equivalently a quantum Cramer-Rao bound~\cite{supp}.  We compute $S_A^{\text{ideal}}$ below.  To visualize how efficiently the information encoded in the scattered light is captured by a particular set of photodetectors and weightings, we construct the differential detection efficiency $\frac{d\eta}{da}$.  We evaluate $d\eta$, the contribution to the overall detection efficiency of a small surface area element $da$ at each position  $\boldsymbol{r}$ in the detection plane, by computing the difference in the overall efficiency with and without the area element $da$ located at $\boldsymbol{r}$.   Define the domain $D'$  equal to $D$ excluding the area element $da$ located at $\boldsymbol{r}$.  Let  $\mathcal{S}'$, $\mathcal{N}'$, and $\eta'$ be defined analogously to Eqns.~\ref{eq:sensitivity}, \ref{eq:noise}, and \ref{eq:eta}, respectively, except that the domain of integration is $D'$ instead of $D$. Then $d\eta=\eta-\eta'$, and we can rewrite the following 
\begin{align*}
\mathcal{S}&=\mathcal{S}'+d\mathcal{S}\\
\mathcal{N}&=\mathcal{N}'+d\mathcal{N}
\end{align*}
where $d\mathcal{S}=2  k\alpha^2f_w\left(\boldsymbol{r}\right)\text{Re}\left[ u_{0}^{\text{out}}\left(\boldsymbol{r}\right)u_{\text{s}}^{\text{out}}\left(\boldsymbol{r}\right)^{*}\right]da$  and $d\mathcal{N}=\alpha^2\left|  u^{\text{out}}_0\left(\boldsymbol{r}\right)\right|^2f_w\left(\boldsymbol{r}\right)^2 da$.  Then we expand $d\eta$ to first order in $da$
\[
d\eta=S_{A}^{\text{ideal}}\frac{\mathcal{S}'^2}{\mathcal{N}'}\left(\frac{2 d\mathcal{S}}{\mathcal{S}'}-\frac{d\mathcal{N}}{\mathcal{N}'} \right)
\]
We can safely approximate $\mathcal{S}'\rightarrow\mathcal{S}$ and $\mathcal{N}'\rightarrow\mathcal{N}$ in the above expression, as they differ only by an infinitesimal contribution.  Then, rearranging, we arrive at
\begin{align*}
\frac{d\eta}{da} =\alpha^2 S_{A}^{\text{ideal}}
&\left(4k\left(\frac{\mathcal{S}}{\mathcal{N}}\right) \text{Re}\left[u_{0}^{\text{out}}(\boldsymbol{r}) u_{\text{s}}^{\text{out}*}(\boldsymbol{r}) f_w(\boldsymbol{r})\right]\right.\\
  & \left.
-\left(\frac{\mathcal{S}}{\mathcal{N}}\right)^2 \left|u_{0}^{\text{out}}(\boldsymbol{r}) f_w(\boldsymbol{r})\right|^{2}\right), \nonumber
\end{align*}
which is the main theoretical result of this paper.

It is straightforward to extend this definition of differential detection efficiency to a generalized area element.  If the system is effectively independent of one transverse spatial coordinate, we can integrate over that coordinate and arrive at $d\eta/dx$.  If our detection plane is in the far field, we can consider $d\eta/d\Omega$, taking our area element to be an infinitesimal patch of solid angle~\cite{supp}.  We can even visualize how detection efficiency is distributed over domains that are not surfaces in real space.  For example, in resolving scattered light into a basis such as Hermite Gaussian or Laguerre Gaussian modes with mode indices $(i,j)$, using a spatial mode sorter (as has been demonstrated for measurements of membranes~\cite{Choi_Pluchar_He_Guha_Wilson_2024} and levitated particles~\cite{Dinter_Roberts_Volz_Schmidt_Laplane_2024}), we can map out $\Delta\eta_{ij}$, the contribution of the $(i,j)$ mode to the over detection efficiency.

To find $S_A^{\text{ideal}}$, the ideal measurement imprecision, an ideal homodyne measurement scheme can be constructed by adding a strong, uniform optical local oscillator (LO) to the output field and performing spatially resolved homodyne detection at every point in the detection plane with independent photodetectors.  We assume that all of the scattered light reaches the detection plane (i.~e.~$\eta_{\text{col}}=1$). Then the field in the detection plane becomes $\alpha u_0^{\text{out}}(\boldsymbol{r})+\alpha kA\,u_{\text{s}}^{\text{out}}(\boldsymbol{r})+\alpha_{\text{LO}}e^{i\phi_{\text{LO}}\left(\boldsymbol{r}\right)}$ , where $\alpha_{\text{LO}}$ is the amplitude of the homodyne LO, and we assume that the polarization of the LO matches the polarization of $u^{\text{out}}_{\text{s}}$. The intensity becomes $I(\boldsymbol{r})\approx\alpha_{\text{LO}}^2+2 kA\alpha \alpha_{\text{LO}}\left|  u_{\text{s}}^{\text{out}}\left(\boldsymbol{r}\right)\right|+2 \alpha \alpha_{\text{LO}}\text{Re}\left[u_0^{\text{out}} e^{i\phi_{\text{LO}}}\right]$, if $\alpha_{\text{LO}}\gg\alpha$ and the LO phase, $\phi_{\text{LO}}$, is chosen to match the phase of the signal field.  If we intend to only look for fluctuations in $A$ (e.~g.~oscillating mechanical systems), as is done in~\cite{Maurer_Gonzalez-Ballestero_Romero-Isart_2023}, then we can ignore the last term, as a constant DC signal.  In this case, we see that the signal amplitude at each location in the detection plane is simply proportional to  $\left|  u_{\text{s}}^{\text{out}}\left(\boldsymbol{r}\right)\right|$ and the noise is dominated by the shot noise on the LO.  Choosing $f_w\left(\boldsymbol{r}\right)=\left|  u_{\text{s}}^{\text{out}}\left(\boldsymbol{r}\right)\right|$ optimally weights each location according to its inverse variance.  Following the same procedure as before, we construct $\mathcal{S}^{\text{ideal}}=2\alpha\alpha_{\text{LO}}k\int_{D}\left|u_{\text{s}}^{\text{out}}\left(\boldsymbol{r}\right)\right|^2da$  and $\mathcal{N}^{\text{ideal}}=\alpha_{\text{LO}}^2\int_{D}\left|u_{\text{s}}^{\text{out}}\left(\boldsymbol{r}\right)\right|^2da$.  Then
\begin{equation}\label{eq:Sideal}
S_{A}^{\text{ideal}}=\frac{\mathcal{N}^{\text{ideal}}}{\left(\mathcal{S}^{\text{ideal}}\right)^2}=\left(4\alpha^2 k^2\int_{D}\left| u_{\text{s}}^{\text{out}}\left(\boldsymbol{r}\right)\right|^2da\right)^{-1}
\end{equation}
We can construct the ideal differential detection efficiency, which coincides with the information radiation pattern when the detection plane is in the far field.  Following the same procedure as before:
\[
\left(\frac{d\eta}{da}\right)^{\text{ideal}}=4\alpha^2 k^2S_{A}^{\text{ideal}}\left|u_{\text{s}}^{\text{out}}\left(\boldsymbol{r}\right)\right|^2
\]


The measurement back action from optical forces can be calculated from Eq.~\ref{eq:Sideal} and the Heisenberg uncertainty bound, 
\begin{equation}\label{eq:Sba}
S_F^{\text{ba}}=\alpha^2 \left(\hbar k\right)^2\int_{D}\left| u_{\text{s}}^{\text{out}}\left(\boldsymbol{r}\right)\right|^2da=\alpha^2 \hbar^2\int_D\left| \frac{du^{\text{out}}}{dA}\right|^2da.
\end{equation}
We can interpret $\bar{N}_{\text{s}}=\alpha^2\int_{D}\left| u_{\text{s}}^{\text{out}}\left(\boldsymbol{r}\right)\right|^2da$ as the rate at which information containing photons are scattered by the system.  Then $\sqrt{\bar{N}_{\text{s}}} \hbar k$ is the noise in the rate of momentum kicks from these signal photons.  The back action can be evaluated in any detection plane $D$, assuming no light is lost in the intervening optics.  Equation~\ref{eq:Sba} is equivalent to integrating the local radiation pressure in the object plane (or a subsequent image plane), and equivalent to integrating the momentum transfer of the signal photons in the far field.   A general strategy to maximize measurement sensitivity then has two parts: find a geometry for the optical probe that maximizes $\bar{N}_{\text{s}}$, and detect all photons in the signal mode $u_{\text{s}}^{\text{out}}\left(\boldsymbol{r}\right)$~\cite{Bouchet_Rotter_Mosk_2021}.

\end{document}


\setcounter{figure}{0}
\renewcommand{\thesection}{S\arabic{section}}  
\renewcommand{\thetable}{S\arabic{table}}  
\renewcommand{\thefigure}{S\arabic{figure}}
\renewcommand{\theequation}{S.\arabic{equation}}

\preprint{APS/123-QED}

\title{Supplemental Materials for:\\Visualizing Measurement Efficiency in Optomechanical Scattering
}

\author{Youssef Tawfik}
\affiliation{Department of Physics and Astronomy, University of Pittsburgh, Pittsburgh, PA, USA}

\author{Shan Hao}%
\altaffiliation[Present address: ]{Apple. Inc}
\affiliation{Department of Physics and Astronomy, University of Pittsburgh, Pittsburgh, PA, USA}

\author{Thomas P. Purdy}%
 \email{Contact Author: tpp9@pitt.edu}
\affiliation{Department of Physics and Astronomy, University of Pittsburgh, Pittsburgh, PA, USA}
\date{\today}
\maketitle

\section{Analysis of Measurement Techniques}

\subsection{QPD Measurement of Membrane Motion}
The vibrational modes of a thin, high-stress rectangular membrane with dimensions $L_x \times L_y$ are sinusoidal out-of-plane modes with $m$ anti-nodes along the $x$ direction and $n$ anti-nodes along the $y$ direction, labeled with mode indices $\left(m,n\right)$. The mode functions are
\[
\psi_{mn}\left(x,y\right)=\sin\left(k_{m}x\right)\cos\left(k_{n}y\right)
\]
for $m$ even and $n$ odd, where $k_m=\pi m/L_x$ and $k_n=\pi n/L_y$ are the mechanical wavenumbers.  We assume the membrane mode frequencies are non-degenerate, so that we can spectrally address individual modes, and so consider only the measurement of a single mode, which is excited to an amplitude $A_{mn}$. We probe the membrane with a Gaussian input beam at normal incidence, centered on the membrane, which has a spatial mode profile
\[
u^{\text{in}}\left(x,y\right)=\sqrt{\frac{1}{\pi w_{0}^{2}}}e^{-\frac{x^{2}}{2w_{0}^{2}}}e^{-\frac{y^{2}}{2w_{0}^{2}}}
\]
where $w_{0}$ is the focused waist of the beam at the membrane surface.  The input beam is in a coherent state with amplitude $\alpha$, wavenumber $k$, and wavelength $\lambda=2\pi/k$.  We assume that the polarization of the beam is constant throughout the experiment.  The input beam is reflected from the undulating surface of the membrane, and the output mode near the membrane surface immediately after reflection is
\begin{align*}
u^{\text{out}}\left(x,y,z=0\right) & =e^{2ikA_{mn}\psi_{mn}\left(x,y\right)}u^{\text{in}}\left(x,y\right)\\
 & \approx u^{\text{in}}\left(x,y\right)+2ikA_{mn}\psi_{mn}\left(x,y\right)u^{\text{in}}\left(x,y\right)\\
 & =u_{0}^{\text{out}}\left(x,y\right)+kAu_{s}^{\text{out}}\left(x,y\right)
\end{align*}
The signal is measured at the detection plane $z=z_{d}$ in the far
field, so we can propagate the output field using the Fraunhofer approximation~\cite{hecht}.
\begin{align*}
u_{0}^{\text{out}}\left(x,y,z_d\right) & =\frac{e^{ikz_{d}}}{i\lambda z_{d}}\int u_{0}^{\text{out}}\left(x',y'\right)e^{ik(xx'+yy')/z_{d}}da'\\
u_{s}^{\text{out}}\left(x,y,z_d\right) & =\frac{e^{ikz_{d}}}{i\lambda z_{d}}\int u_{s}^{\text{out}}\left(x',y'\right)e^{ik(xx'+yy')/z_{d}}da'
\end{align*}
In this case, $u_0^{\text{out}}\left(\boldsymbol{r}\right)$ is a Gaussian beam of waist $w_d\approx z_d/k w_0$, and  $u_{\text{s}}^{\text{out}}\left(\boldsymbol{r}\right)$ is proportional to the spatial Fourier transform of $\psi_{mn} u^{\text{in}}$.

For the QPD measurement, the weight function is $f_{w}\left(x,y\right)$
is $1$ for $x>0$ and $-1$ for $x<0$. The signal and noise integrals defined in the main text evaluate to
\begin{align*}
\mathcal{N} & =\int_{D}\left|\alpha u_{0}^{\text{out}}\left(\boldsymbol{r}\right)\right|^{2}f_{w}\left(\boldsymbol{r}\right)^{2}da=\left|\alpha\right|^{2}\\
\mathcal{S} & =2\left|\alpha\right|^{2}k\int_{D}\text{Re}\left[u_{0}^{\text{out}*}\left(\boldsymbol{r}\right)u_{s}^{\text{out}}\left(\boldsymbol{r}\right)\right]f_{w}\left(\boldsymbol{r}\right)da\\
 & =-4\left|\alpha\right|^{2}k\,\text{erf}\left(k_{m}w_{0}/2\right)e^{-\frac{1}{4}k_{m}^{2}w_{0}^{2}}e^{-\frac{1}{4}k_{n}^{2}w_{0}^{2}}\\
\mathcal{I} & =4k^{2}\int_D\left|\alpha u_{s}^{\text{out}}\left(\boldsymbol{r}\right)\right|^{2}da\\
 & =4\left|\alpha\right|^{2}k^{2}\left(1-e^{-k_{m}^{2}w_{0}^{2}}\right)\left(1+e^{-k_{n}^{2}w_{0}^{2}}\right)
\end{align*}
 From these quantities, we derive the imprecision noise $S_{A_{mn}}^{\text{imp}}=\mathcal{N}/\mathcal{S}^{2}$,
ideal imprecision noise $S_{A_{mn}}^{\text{ideal}}=1/\mathcal{I}$, and the detection efficiency $\eta=S_{A_{mn}}^{\text{ideal}}/S_{A_{mn}}^{\text{imp}}$.  

For the case of a partially blocked QPD, we numerically evaluate the above integrals using the weight function 
\[f_w\left(x,y\right)=\left\{
\begin{aligned}
1&,\, x>B/2  \\
-1&,\, x<-B/2  \\
0&,\, \text{otherwise}
\end{aligned}
\right.\]
where $B$ is the width of the blocked region.  

\subsubsection{Optical Lever Limit}
In the limit where $k_m w_0\ll 1$ and $k_n w_0\ll1$, the membrane, locally, looks like a flat tilting surface, the optical lever limit.  We can approximate $\psi_{mn}\left(x,y\right)\approx k_m x$. The far field modes simplify to $u_0^{\text{out}}\left(\boldsymbol{r}\right)=-ie^{i k z_d}\sqrt{\frac{1}{\pi w_d^2}} e^{-\frac{x^2+y^2}{2 w_d^2}}$ and $u_{\text{s}}^{\text{out}}\left(\boldsymbol{r}\right)=-2  k_m w_0\frac{x}{w_d} u_0^{\text{out}}\left(\boldsymbol{r}\right)$.  The signal mode is of the form of $HG_{1,0}\left(x,y\right)$, a first order Hermite Gaussian mode.  The ideal imprecision noise becomes
\[
S_{A_{mn}}^{\text{Ideal}}=\frac{1}{8k^{2}k_{m}^{2}w_{0}^{2}}\frac{1}{\alpha^{2}}
\]
By integrating the mode functions over their $y$ coordinate, we can compute the one-dimensional ideal DDE
\[
\left(\frac{d\eta}{dx}\right)^{\text{ideal}}=\frac{2}{\sqrt{\pi w_d^2}}\frac{x^2}{w_d^2}e^{-\frac{x^2}{w_d^2}}
\]
The sensitivity for detection with a QPD simplifies to $\mathcal{S}=-4 \left| \alpha\right|^2 k k_m w_0/\sqrt{\pi}$, and then
\[\eta=\frac{2}{\pi}\]
We find that for a partially blocked QPD in the optical lever limit
\[
\eta=\frac{2}{\pi}\frac{e^{-\frac{B^2}{2 w_d^2}}}{2\left(1-\text{erf}\left(\frac{B}{2 w_d}\right)\right)}
\]
This function is maximized when $B=0.87 w_d$ where $\eta=0.81$
 and about 46\% of the optical power is absorbed by the block.  (For higher order modes, $B$ is optimized at larger values.) 

 \subsubsection{Linear Weighting Function}
 In the optical lever limit, instead of using the QPD weighting function, we can choose a linear weighting function (e.~g.~$f_w(x,y)=\frac{x}{w_d}$).  Then 
\[
\mathcal{N}=\alpha^{2}/2,\quad\mathcal{S}=-2\alpha^{2}k k_{m} w_0
\]
We find that $\eta=1$, and the DDE is equal to its ideal version.  Such a weighting function could be implemented in principle with a high speed, high dynamic range camera by finding the centroid of the intensity distribution on the camera.  Alternatively, a lateral effect position sensing photodiode can implement this weighting function.  These devices consist of a large area photodiode backed by a resistive layer, through which the photocurrent must travel before reaching the electrodes.  Such devices are able to outperform QPDs at the same optical power, if both reach the shot noise limit, as shown in Ref.~\cite{Fradgley_French_Rushton_Dieudonné_Harrison_Beckey_Miao_Gill_Petrov_Boyer_2022}.

\begin{figure}  
  \centering
  \includegraphics[width=\textwidth]{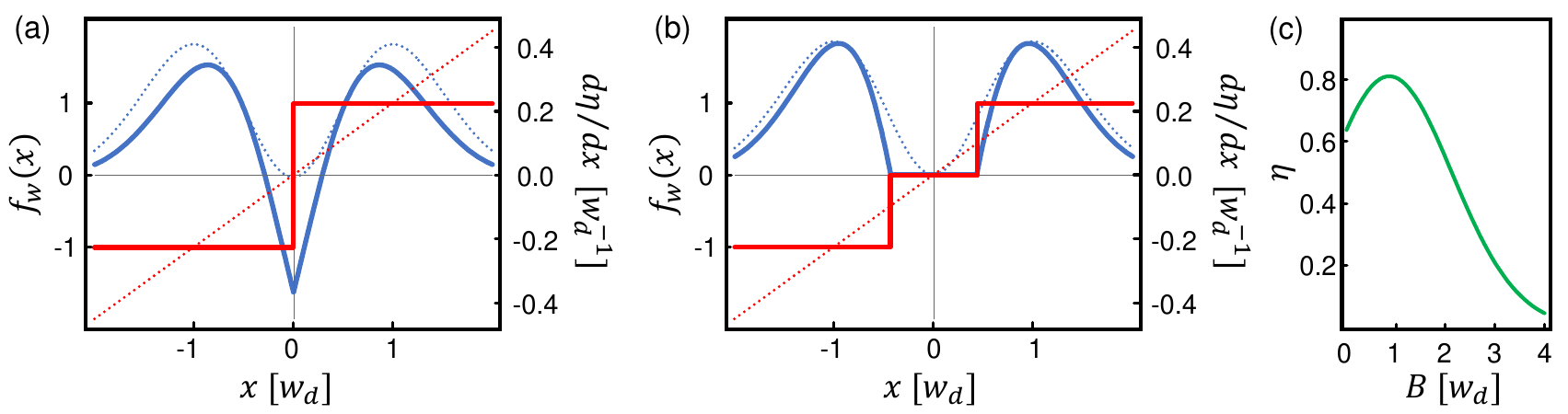}\label{fig:optical_lever}
  \caption{QPD measurement in the optical lever limit.  (a) Standard QPD detection.  Weighting function, $f_w$ (red), ideal weighting function (dotted red), DDE (blue), ideal DDE (dotted blue) (b) Partially blocked QPD detection.  Trace colorings are the same as in panel (a).  The weight function in (b), as compared to (a), more closely approximates the ideal linear weighting function, which results in the DDE more closely approximating the ideal one. (c) Detection efficiency as a function of block width.
  }
  \label{fig:optical_lever_limit}
\end{figure}

 \subsubsection{Single-Mode Interferometry}
For comparison, we calculate the noise in a single-mode interferometer detecting the motion of the membrane.  We assume that the input laser is focused to an anti-node of a membrane mode, so that we can approximate $\psi_{mn}(x,y)=1$.  Light is reflected from the membrane and then detected with a mode matched optical LO, with phase set to detect the phase quadrature of the output light.  This scheme is equivalent to an unbalanced Mach-Zehnder interferometer.  The output field becomes $\alpha u_0^{\text{out}} \left(1+2 i k A\right)+i\alpha_{\text{LO}}u_0^{\text{out}}$.   We assume $\alpha_{\text{LO}}\gg\alpha$. Then $\mathcal{S}=4\alpha \alpha_{\text{LO}}k$ and $\mathcal{N}=\alpha_{\text{LO}}^2$, from which $S_{A_{mn}}^{\text{interferometer}}=\frac{1}{16 \alpha^2 k^2}$ and $S_{F_{mn}}^{\text{ba}}=\alpha^2\left(2\hbar k\right)^2$.  As expected, this detection scheme saturates the Heisenberg uncertainty limit with $\eta=1$.  For a given level of optical power, focusing all the light where the amplitude of motion is maximal generates the largest optical force on the membrane, and thus also yields the lowest possible measurement noise floor.  We can benchmark our QPD based measurement schemes against this single mode interferometry to assess not only the detection efficiency, but also the absolute level of the measurement noise floor.
 

\subsection{Phase Contrast Imaging of Membrane Modes}
Because information about membrane motion is imprinted as a spatially dependent phase on the reflected light, it is natural to consider detection via phase contrast imaging.  Here, light reflected from a membrane is imaged onto the detection plane, while $u_0^{\text{out}}$ receives an extra $\pi/2$ phase shift via an intermediate manipulation in a Fourier plane~\cite{hecht}.  
\[u^{\text{out}}\left(x,y,z_d\right)=i u^{\text{in}}\left(x,y\right)+i2 k \Psi\left(x,y\right)u^{\text{in}}\left(x,y \right)
\]
For membrane modes where $k_m\ll k$. we assume that we can easily collect all of the diffracted light with an a modest NA lens, so $\eta_{\text{col}}=1$.  We identify $u^{\text{out}}_{\text{s}}=2i\psi_{mn}u^{\text{in}}$, for the measurement of a particular mechanical mode.  The phase contrast imaging process turns phase information in the device plane into intensity variations in the image plane, with an example intensity distribution shown in Fig.~\ref{fig:phase_contrast_imaging}(a), for the (14,1) mechanical mode.  The mechanical mode shape is superposed on top of the Gaussian envelope of the input optical beam.
\[I\left(x,y \right)=\alpha^2\left|u^{\text{in}}\left(x,y\right)\right|^2 \left(1+4 k A_{mn}\psi_{mn}\left(x,y\right)\right)
\]
The ideal imprecision noise for phase contrast imaging is the same as for detection in the far field, as no information is lost or created in the free space propagation of the scattered light~\cite{Hupfl_Russo_Rachbauer_Bouchet_Lu_Kuhl_Rotter_2024}.  The ideal DDE is simply $16 k^2 \alpha^2 S^{\text{ideal}}_{A_{mn}}\left| u^{\text{in}}\right|^2\left(\psi_{mn}\right)^2$.  An ideal detection scheme can be implemented with a high pixel resolution camera by applying a weighting function to each pixel $f_w\left(x,y\right)=\psi_{mn}\left(x,y\right)$.  However, for the case of measuring the motion of a $\sim$MHz frequency mode with appreciable optical power, most cameras will not have the necessary frame rate or dynamic range.  For modes with $n=1$ and $k_n w_0\ll 1$, as discussed previously, the problem becomes effectively one dimensional, and we can integrate over $y$, dealing only with the spatial variations along $x$.  A linear photodiode array with 2 elements per mechanical period will minimally satisfy the spatial Nyquist sampling rate, while operating with high bandwidth and dynamic range.  We assume that we can adjust the magnification of our imaging system such that the pitch of the photodiode array is equal to half the mechanical wavelength in the image plane. Weighting these detectors with alternating values of $\pm1$ is equivalent to $f_w=\text{signum}(\psi_{m1})$.  In the limit $k_m w_0 \gg 1$
\[\mathcal{S}=8 \alpha^2 k/\pi;\,\,  \mathcal{N}=\alpha^2;\,\, S_{A_{m1}}^{\text{ideal}}=\left(8\alpha^2 k^2\right)^{-1},
\]
yielding $\eta\approx 8/\pi^2=0.81$, independent of $m$.  As seen in Fig.~\ref{fig:phase_contrast_imaging}(b), the DDE for this configuration is negative in the regions near the border of two detectors, where $\psi_{m1}$ is small.  We find that adding gaps between the detectors (as already exist in commercially available arrays) can increase the detection efficiency to $\eta=0.92$.

 \begin{figure}  
  \centering
  \includegraphics[width=0.75\textwidth]{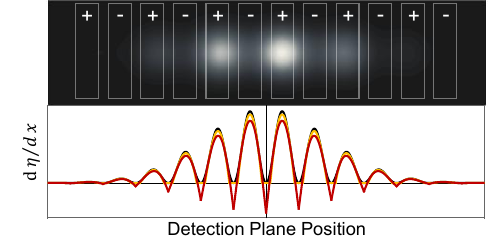}
  \caption{
  Differential detection efficiency of Phase Contrast Imaging. (Top) Simulation of phase contrast image of the (14,1) membrane mode with a linear array of photodetectors depicted as white outlines. (Bottom) Differential detection efficiency with no gap between photodetectors (Red), with an optimized gap between the photodetectors (Blue), and the ideal DDE (Black).
  }
  \label{fig:phase_contrast_imaging}
\end{figure}

\begin{figure}  
  \centering
  \includegraphics[width=\textwidth]{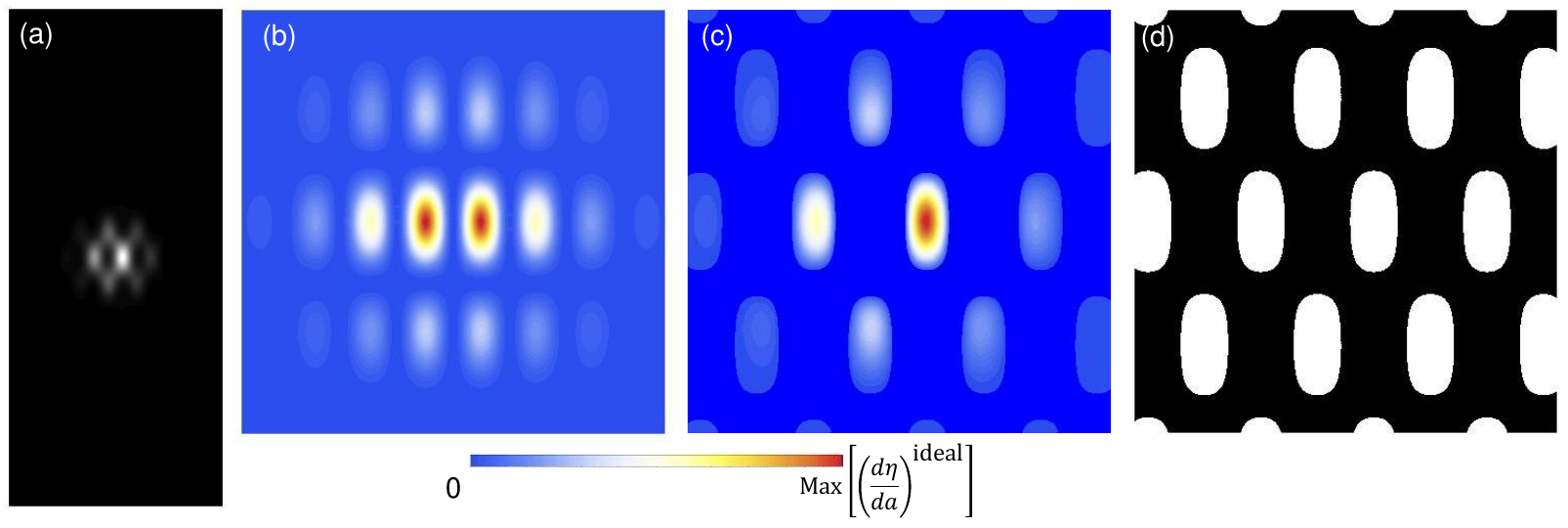}
  \caption{Phase contrast imaging 2D. (14,15) mode.  (a) Intensity distribution in the image plane of a 1.5~mm $\times$ 3.5~mm membrane with (14,15) mode excited. Here, $w0=200$~\textmu m. (b) Ideal DDE zoomed into a 400 \textmu m $\times$ 400 \textmu m field of view. (c) DDE over same field of view as (b) for light blocking mask shown in panel (d). (d) Light blocking mask for single photodetector phase contrast sensing. Black is opaque and white is transparent.
  }
  \label{fig:phase_contrast_imaging_2D}
\end{figure}

Phase contrast imaging is also useful to measure higher order mechanical modes with large values of both $m$ and $n$, where quadrant photodetection would typically yield poor results.  Here, rather than using a 2 dimensional array of photodiodes (whose complexity would start to approach that of a conventional camera), we suggest a detection scheme employing only a single photodetector.  We can implement the weight function
\[f_w\left(x,y\right)=\left\{
\begin{aligned}
& 1 ,\psi_{mn}\left(x,y\right)> \psi_{\text{threshold}}   \\
& 0, \text{otherwise}
\end{aligned}
\right.\]
by partially blocking the photodetector with a mask patterned as shown in Fig.~\ref{fig:phase_contrast_imaging_2D}(d).   Here, the mask is opaque where $f_w=0$  and transparent where $f_w=1$.  All of the light transmitted through the mask is collected on a single photodetector.  We numerically optimize the detection threshold value, $\psi_{\text{threshold}}$, to maximize the detection efficiency. We find that $\psi_{\text{threshold}}\approx0.29$ yields $\eta=0.43$.  This result is again independent of $m$ and $n$ in the diffraction grating limit.

\subsection{Dipolar Scatterer}

We consider the measurement of the center of mass coordinates $\boldsymbol{r}_{0}=(x_0,y_0,z_0)$ of a sub-wavelength size particle via light scattering.  The measurement setup for this scheme is illustrated in Fig.~1(b) of the main text.  A high numerical aperture input lens is filled with a plane wave, and creates a highly focused field centered at the origin.  The output field is the sum of the unscattered input field and the field radiated by the particle, which we approximate as a point dipole.  This output is collected by a collection lens with a  numerical aperture $\text{NA}_{\text{cl}}$, which we assume to be equal to or smaller than the numerical aperture of the input lens.  To compute the far field distribution, we follow the calculations presented in Ref.~\cite{Tebbenjohanns_Frimmer_Novotny_2019}.  The total electric field incident on the input lens is a plane wave of amplitude $\boldsymbol{E}_{\text{inc}}$, which is propagating in the $z$ direction and is polarized in the $x$ direction. This field is transformed by the focusing lens into a focused field. The far field distribution of the unscattered light becomes $\boldsymbol{E}_\infty$. We define the far-field, vector-valued mode function $\boldsymbol{u}_\infty$, in spherical polar coordinates, $\theta$ and $\varphi$ as

\[
\boldsymbol{u}_{\infty}\left(\theta, \varphi\right)=\frac{1}{\sqrt{\pi}}\left(-\sin\varphi\,\hat{\varphi}+\cos\varphi\,\hat{\theta}\right)\sqrt{\cos\theta}
\]
Such far-field mode functions are related to the electric field distribution by $\alpha\boldsymbol{u}/r=\sqrt{\epsilon_{0}c}\boldsymbol{E}$.  We assume that the $\textbf{u}_\infty$ distribution is truncated by the aperture of the collection lens before reaching the detector.  Then $\textbf{u}_\infty$ is normalized such that the rate of photons passing through the collection lens is $\left|\alpha_0\right|^2 \text{NA}_\text{cl}^2$. We must explicitly track the polarization of the fields, because in this measurement the interfering fields no longer have the same polarization (unlike in the previous examples where light is reflected from a nearly flat membrane at normal incidence and a uniform polarization is assumed). 

The dipolar  scatterer experiences the time harmonic focal field at position $\boldsymbol{r}_{0}$ and radiates with a mode pattern
\begin{align*}
\boldsymbol{u}_{\text{rad}}\left(\boldsymbol{r}\right) & =\boldsymbol{u}_{\text{dip}}\left(\boldsymbol{r}\right)e^{-ik\left(\hat{\boldsymbol{r}}\cdot\boldsymbol{r}_{0}\right)}e^{ikA_{G}z_{0}}\\
 & \approx\boldsymbol{u}_{\text{dip}}\left(\boldsymbol{r}\right)\left(1-ik\left(\hat{\boldsymbol{r}}\cdot\boldsymbol{r}_{0}-A_{G}z_{0}\right)\right)
\end{align*}
\[
\boldsymbol{u}_{\text{dip}}\left(\boldsymbol{r}\right)=\sqrt{\frac{3}{8\pi}}\left(\hat{r}\left(\hat{r}\cdot\hat{x}\right)-\hat{x}\right)
\]
where $A_{G}$ is a Gouy phase correction \cite{nanooptics}, for a Gaussian beam this correction is $1-k z_R$ where $z_R$ is the Rayleigh range. $\boldsymbol{u}_{\text{dip}}\left(\boldsymbol{r}\right)$
is the radiation pattern from a dipole radiating at the origin.  The total scattering from the particle is $\left|\alpha_{\text{dip}}\right|^{2}$, which depends on the details of the composition of the particle.  The field near the focus which drives the scatterer is $\pi/2$ out of phase with $\boldsymbol{u}_\infty$, due to the Gouy phase of propagation. Then, the dipole radiation field is $\pi/2$ out of phase with $\boldsymbol{u}_\infty$. The input and radiated field interfere on the reference sphere of the collection lens. 
\begin{align*}
 \alpha_0 \boldsymbol{u}_{\infty}\left(\boldsymbol{r}\right)+i\alpha_\text{dip} \boldsymbol{u}_{\text{rad}}\left(\boldsymbol{r}\right)& =\left(\alpha_0\boldsymbol{u}_{\infty}+i\alpha_\text{dip}\boldsymbol{u}_{\text{dip}}\right)+\alpha_\text{dip}k\left(\hat{\boldsymbol{r}}\cdot\boldsymbol{r}_{0}-Az_{0}\right)\boldsymbol{u}_{\text{dip}}\\
 & =\alpha_0\boldsymbol{u}_{0}^{\text{out}}+\alpha_\text{dip}k\left(\hat{\boldsymbol{r}}\cdot\boldsymbol{r}_{0}-Az_{0}\right)\boldsymbol{u}_{s}^{\text{out}}\\
 & =\alpha_0\boldsymbol{u}_{0}^{\text{out}}+\alpha_\text{dip}\left(kx_{0}\boldsymbol{u}_{s,x_{0}}^{\text{out}}+ky_{0}\boldsymbol{u}_{s,y_{0}}^{\text{out}}+kz_{0}\boldsymbol{u}_{s,z_{0}}^{\text{out}}\right)
\end{align*}
where $\alpha_0$ and $\alpha_\text{dip}$ are real valued and we have assumed $\alpha_0 \gg \alpha_\text{dip}$. 

Using the above form of the output distributions we can generate the signal and
noise using a general weight function as in the main text, except that since the
polarization of the two fields is no longer the same, we generalize
the inner products of the scalar mode functions $u$ to include dot products between polarization vectors. Since we use non-dimensionalized, far-field mode functions, our integrals change from area to solid angle.  We can employ the formulas from the main text by substituting $\left| \alpha \frac{\boldsymbol{u}}{r} \right|^2da=\left|\alpha \boldsymbol{u}\right|^2d\Omega$.
\begin{align*}
\mathcal{N} & =\left|\alpha_{0}\right|^{2}\int_{D_c}\left|\boldsymbol{u}_{0}^{\text{out}}\left(\boldsymbol{r}\right)\right|^{2}f_{w}\left(\boldsymbol{r}\right)^{2}d\Omega\\
 & =\frac{\left|\alpha_{0}\right|^{2}}{\pi}\int_{D_c}\cos\theta f_{w}\left(\boldsymbol{r}\right)^{2}d\Omega
\end{align*}
where $D_c$ is the domain of the solid angle subtended by the collection lens.  For a standard QPD this evaluates to
\[
\mathcal{N}=\left|\alpha_{0}\right|^{2}\text{NA}_{\text{cl}}^2
\]
The sensitivity of the measurement of $x_{0}$ and $y_{0}$ will be
given by,
\begin{align*}
\mathcal{S}_{x_{0}} & =2k\alpha_{0}\alpha_{\text{dip}}\int_{D_c}\text{Re}\left[\boldsymbol{u}_{0}^{\text{out}*}\cdot\boldsymbol{u}_{s,x_{0}}^{\text{out}}\right]f_w\left(\boldsymbol{r}\right)d\Omega\\
 & =-\sqrt{\frac{3}{2}}\frac{k}{\pi}\alpha_{0}\alpha_{\text{dip}}\int_{D_c}\left(\sin^{2}\varphi\,+\cos^{2}\varphi\cos\theta\right)\sin\theta\cos\varphi\sqrt{\cos\theta}f_{w}\left(\boldsymbol{r}\right)d\Omega
\end{align*}

\begin{align*}
\mathcal{S}_{y_{0}} & =2k\alpha_{0}\alpha_{\text{dip}}\int_{D_c}\text{Re}\left[\boldsymbol{u}_{0}^{\text{out}*}\cdot\boldsymbol{u}_{s,y_{0}}^{\text{out}}\right]f_{w}\left(\boldsymbol{r}\right)d\Omega\\
 & =-\sqrt{\frac{3}{2}}\frac{k}{\pi}\alpha_{0}\alpha_{\text{dip}}\int_{D_c}\left(\sin^{2}\varphi\,+\cos^{2}\varphi\cos\theta\right)\sin\theta\sin\varphi\sqrt{\cos\theta}f_{w}\left(\boldsymbol{r}\right)d\Omega
\end{align*}
As noted in Ref.~\cite{Tebbenjohanns_Frimmer_Novotny_2019}, most of the information about $z_0$ is contained in the back-scattered field, so we do not consider detection of motion along the optical axis with our forward scattering detection scheme.  The ideal measurement signal is given by,
\begin{align*}
\mathcal{I}_{x_{0}} & =4k^{2}\left|\alpha_{\text{dip}}\right|^{2}\int\left|\boldsymbol{u}_{s,x_{0}}^{\text{out}}\left(\boldsymbol{r}\right)\right|^{2}d\Omega,\\
 & =\frac{4}{5}k^{2}\left|\alpha_{\text{dip}}\right|^{2}
\end{align*}
\begin{align*}
\mathcal{I}_{y_{0}} & =4k^{2}\left|\alpha_{\text{dip}}\right|^{2}\int\left|\boldsymbol{u}_{s,y_{0}}^{\text{out}}\left(\boldsymbol{r}\right)\right|^{2}d\Omega\\
 & =\frac{8}{5}k^{2}\left|\alpha_{\text{dip}}\right|^{2}
\end{align*}
From these quantities, we can calculate the detection efficiency of
this measurement $\eta_{x_{0}}=S_{x_{0}}^{\text{ideal}}/S_{x_{0}}^{\text{imp}}$
and the differential detection efficiency $d\eta/d\Omega$.

The information radiation pattern for a dipolar scatterer is~\cite{Tebbenjohanns_Frimmer_Novotny_2019}
\[
\frac{d\eta_{x_{0}}^{\text{ideal}}}{d\Omega}=\frac{15}{8\pi}\left(1-\sin^{2}\theta\cos^{2}\varphi\right)\sin^{2}\theta\cos^{2}\varphi
\]
\[
\frac{d\eta_{y_{0}}^{\text{ideal}}}{d\Omega}=\frac{15}{16\pi}\left(1-\sin^{2}\theta\cos^{2}\varphi\right)\sin^{2}\theta\sin^{2}\varphi
\]

 \begin{figure}  
  \centering
  \includegraphics[width=0.75\textwidth]{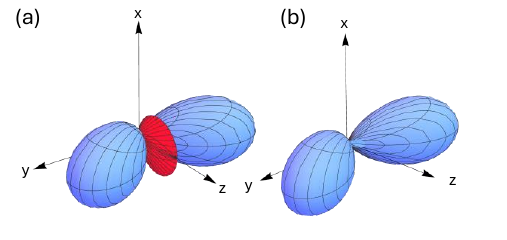}
  \caption{
  Differential detection efficiency of a dipolar scatterer $d\eta_y/d\Omega$ for a collection lens with $\text{NA}_\text{cl}  = 1$, where blue represents positive values of the DDE and red represents negative values. (a) Standard QPD, (b) Optimally Blocked QPD.
  }
  \label{fig:phase_contrast_imaging}
\end{figure}

\section{Fisher Information}

We can construct the ideal measurement noise floor for an optomechanical scattering experiment by computing the quantum Fisher information (QFI) contained in the output light~\cite{Bouchet_Rotter_Mosk_2021,Hupfl_Russo_Rachbauer_Bouchet_Lu_Kuhl_Rotter_2024}.  To simplify the calculation, we can break the signal field into two orthogonal components~\cite{Pluchar_He_Manley_Deshler_Guha_Wilson_2025}. We define $u^{\text{out}}_{s\perp}$ as the component of the signal field orthogonal to the stationary field, where $\int u_{0}^{\text{out}}\left(\boldsymbol{r}\right)^*u_{\text{s}\perp}^{\text{out}}\left(\boldsymbol{r}\right)da=0 $.  Additionally, the stationary output field can acquire a phase shift $\phi_I(kA)$, as in a traditional single-mode interferometer.  Then $u^{\text{out}}\approx u_0^{\text{out}}e^{i\phi_{I}(kA)}+k A u^{\text{out}}_{s\perp} \approx u_0^{\text{out}}+k A\left( i \phi_I u_{0}^{\text{out}}+\, u_{\text{s }\perp}^{\text{out}}\right)$, where $\phi_{I}$ is the interferometric sensitivity, and we have expanded this expression to first order in $kA$. We then identify $u^{\text{out}}_{\text{s}}= i \phi_I u_{0}^{\text{out}}+\, u_{\text{s }\perp}^{\text{out}}$.  The amplitude of the stationary field, $u_0^{\text{out}}$, remains unchanged to first order in $k A$.

We write the quantum state of the output light as $\Ket{\psi^{\text{out}}}=\ket{\alpha_0}_0\otimes \ket{\alpha_{\perp}}_{\perp}$, where the first ket represents a coherent state with amplitude $\alpha_0$ in the $u_0^{\text{out}}\left(\boldsymbol{r}\right)$ spatial mode, and the second ket represents a coherent state with amplitude $\alpha_{\perp}$ in  the $\tilde{u}^{\text{out}}_{\text{s}\perp}\left(\boldsymbol{r}\right)$ spatial mode.  Here, $\tilde{u}^{\text{out}}_{\text{s}\perp}\left(\boldsymbol{r}\right)=\frac{u^{\text{out}}_{\text{s}\perp}\left(\boldsymbol{r}\right)}{n_{\perp}}$ is a normalized version of the orthogonal component of the signal mode, where $n_{\perp}=\int_D\left|u^{\text{out}}_{\text{s}\perp} \right|^2da$.  Then, the output state for $A=0$ is  $\Ket{\psi^{\text{out}}_0}=\ket{\alpha}_0\otimes \ket{0}_{\perp}$, and  $\Ket{\psi^{\text{out}}}=\ket{\alpha\left(1+i\phi_{I}k A \right)}_0\otimes \ket{\alpha k A \sqrt{n_{\perp}}}_{\perp}$.  A unitary transformation between $\Ket{\psi^{\text{out}}_0}$  and $\Ket{\psi^{\text{out}}}$ corresponds to a phase shift for the $u_0^{\text{out}}$ mode and a displacement for the $\tilde{u}^{\text{out}}_{\text{s}\perp}$ mode.
\[
\ket{\alpha\left(1+i\phi_{I}k A \right)}_0\otimes \ket{\alpha k A \sqrt{n_{\perp}}}_{\perp}=\hat{\Phi}_0(A)\otimes\hat{D}_{\perp}(A)\ket{\alpha}_0\otimes\ket{0}_{\perp}
\]
where
\begin{align*}
\hat{\Phi}_0(A)&=e^{-ikA\phi_I\hat{a}^{\dagger}_0\hat{a}_0}=e^{i A\hat{H}_0}\\
\hat{D}_{\perp}(A)&=e^{k A \sqrt{n_{\perp}}\left(\alpha \hat{a}_{\perp}^{\dagger}-\alpha^{*}\hat{a}_{\perp}\right)}=e^{i A\hat{H}_{\perp}}
\end{align*}
with $\hat{a}_0$ and $\hat{a}_{\perp}$ being the mode annihilation operators.  The generators of these transformations are $\hat{H}_0=k\phi_I\hat{a}^{\dagger}_0\hat{a}_0$ and $\hat{H}_{\perp}=-i k \sqrt{n_{\perp}}\left(\alpha \hat{a}^{\dagger}_{\perp}-\alpha^{*}\hat{a}_{\perp}  \right)$, so that $\hat{\Phi}_0(A)\otimes \hat{D}_{\perp}(A)=e^{i A (\hat{H}_0+\hat{H}_{\perp})}$.  For the case of pure input states, such as coherent states, the QFI is given by
\[
\mathcal{F}_Q=4\sigma^2_{\hat{H}_0 + \hat{H}_{\perp}}=4\left(\sigma^2_{\hat{H}_0}+\sigma^2_{\hat{H}_{\perp}}\right)
\]
assuming that the two modes are orthogonal.  These variances can be evaluated for coherent states,  $\sigma^2_{\hat{H}_0}=k^2\phi_I^2 \left| \alpha \right|^2$ and $\sigma^2_{\hat{H}_{\perp}}=k^2\left| \alpha \right|^2 n_{\perp}$.  Then the QFI evaluates to
\begin{align*}
\mathcal{F}_Q=&4 k^2\phi_I^2\left| \alpha \right|^2+4 k^2 \left| \alpha \right|^2 \int_D \left| u^{\text{out}}_{\text{s}\perp}\right|^2 da\\
=&4 k^2 \left|\alpha\right|^2\int_D \left|u^{\text{out}}_{\text{s}}\right|^2da\\
=&4 \left|\alpha\right|^2\int_D \left| \frac{d u^{\text{out}}}{dA} \right|^2da
\end{align*}
where we use the definition $u^{\text{out}}_{\text{s}}= i \phi_I u_{0}^{\text{out}}+\, u_{\text{s }\perp}^{\text{out}}=\frac{1}{k}\frac{d u^{\text{out}}}{dA}$ from the main text and assume that $u^{\text{out}}_0$ is normalized.  We can alternatively arrive at the same conclusion, if we use the more general form for the QFI, $\mathcal{F}_Q=4\Braket{\frac{d \psi^{\text{out}}}{dA}|\frac{d \psi^{\text{out}}}{dA}}-\left|\Braket{\frac{d\psi^{\text{out}}}{dA}|\psi^{\text{out}}} \right|^2$, evaluating this formula by expanding the coherent states in the Fock state basis~\cite{Bouchet_Rotter_Mosk_2021}.

The quantum Cramer-Rao bound for our measurement is
\[
\sigma_A^2\geq\frac{1}{\tau \mathcal{F}_Q}
\]
for an integration time of $\tau$. Saturating this bound yields an ideal measurement noise floor equal to that given in the main text.

\section{Experimental Details}
\begin{figure}   
  \centering
  \includegraphics{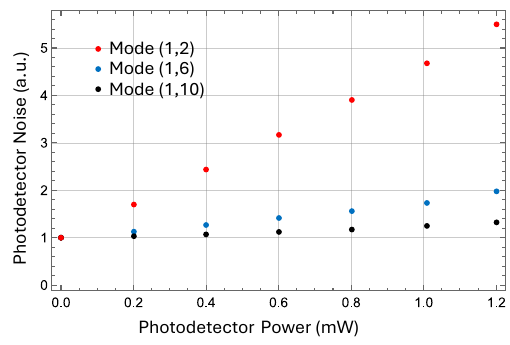}
  \caption{ Optical shot noise calibration.  Measured noise floor as a function of optical power incident on the QPD, near several mechanical modes.  Spectra are normalized to the dark noise floor.  Because the QPD has a frequency dependent gain and dark noise floor, the slopes of these curves differ.  All fit well to a linear trend.}
  \label{fig:Shotnoise}
\end{figure}

Our laser source for this experiment (NKT Photonics, Koheras ADJUSTIK fiber laser) has intensity fluctuations above the shot noise limit for the frequencies and powers used in this study.  However, because we operate our QPD in a balanced configuration, we can achieve shot-noise-limited performance.  The QPD (Thorlabs, PQD30C) has a frequency dependent gain, dark noise floor, and common mode rejection ratio over the range of frequencies of our mechanical modes.  We verified that the photodetector was shot noise limited in frequency bands near the mechanical modes of interest.  We varied the incident power on the QPD by linearly attenuating the beam with an optical waveplate and polarizer, and then recorded the average power spectrum as a function of the incident optical power incident.  In frequency bands near each mechanical mode, we plot the noise floor as a function of incident power, as shown in Fig.~\ref{fig:Shotnoise}.  On top of a constant dark noise contribution, the noise increases linearly with the incident power up to at least a power of 1~mW.  This linear scaling is indicative of shot noise. All measurements in the main text were taken with a power of less than 1 mW incident on the photodetector, and the average dark noise floor is subtracted from spectra before further analysis.

\begin{figure} 
  \centering
  \includegraphics{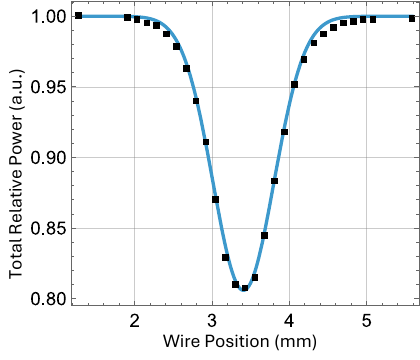}
  \caption{Optical mode profile in detection plane.  Total power detected on the QPD (black squares) is measured as a thin wire is scanned in front of the detector.  A Gaussian fit (blue curve) is used to extract the waist $w_d$.  The actual optical mode profile is one minus the measured values.}
  \label{fig:waistfitting}
\end{figure}

The optical beam profile at the detection plane, $w_d$, is measured using the same experimental configuration as shown in Fig.~2(a) of the main text used to measure the DDE. The thin wire is moved laterally across the front of the photodetector, and the total power summed over both halves the QPD is recorded at each location $x$, as shown in Fig.~\ref{fig:waistfitting}.  We fit this profile to a Gaussian dip to extract $w_d$.  Deviations from a perfect Gaussian profile are visible in the wings of the distribution, which cause deviations of the measured DDE from those calculated assuming a Gaussian mode shape.

We also note that because the diameter of our thin wire is only a few times smaller than the beam waist, the expected DDE is modified from the profile calculated assuming an infinitesimal width block.  To facilitate comparison between experiment and theory, the expected DDEs plotted in Fig.~2 of the main text are convolved with a finite width block function.  The DDEs presented in Fig.~\ref{fig:optical_lever_limit} of this supplement assume an infinitesimal block.

\bibliography{apssamp}